\documentclass[twocolumn]{aastex62}
\usepackage[utf8]{inputenc}
\usepackage{color, colortbl}
\usepackage{soul}
\usepackage{savesym}

\usepackage{url}
\usepackage{hyperref}
\usepackage{amsmath, amssymb}
\usepackage{natbib}
\usepackage{ulem}
\usepackage{bm}

\newcommand{\sne}{SNe\,Ia}
\newcommand{\sn}{SN\,Ia}
\usepackage{savesym}
\savesymbol{tablenum}
\usepackage{siunitx}
\restoresymbol{SIX}{tablenum}
\DeclareSIUnit\parsec{pc}
\DeclareSIUnit\mag{mag}
\DeclareSIUnit\days{days}
\newcommand{\neff}{N_{\text{eff}}}
\newcounter{magicrownumbers}
\newcommand\rownumber{\stepcounter{magicrownumbers}\arabic{magicrownumbers}}

\newcommand{\Einstein}{NASA Einstein Fellow}
\newcommand{\UCSC}{Department of Astronomy and Astrophysics, University of California, Santa Cruz, CA 95064, USA}
\newcommand{\JHU}{Department of Physics and Astronomy, The Johns Hopkins University, Baltimore, MD 21218.}
\newcommand{\STScI}{Space Telescope Science Institute, Baltimore, MD 21218.}
\newcommand{\Duke}{Department of Physics, Duke University, Durham North Carolina 27708, USA}
\newcommand{\Harvard}{Harvard-Smithsonian Center for Astrophysics, 60 Garden Street, Cambridge, MA 02138, USA}
\newcommand{\Kavli}{University of Chicago, Kavli Institute for Cosmological Physics, Chicago, IL, USA.}
\newcommand{\Rutgers}{Department of Physics and Astronomy, Rutgers, The State University of New Jersey, 136 Frelinghuysen Road, Piscataway, NJ 08854, USA}
\newcommand{\USC}{Department of Physics and Astronomy, University of South Carolina, 712 Main Street, Columbia, SC 29208, USA}

\newcommand{\NCUG}{Graduate Institute of Astronomy, National Central University, 300 Zhongda Road, Zhongli, Taoyuan 32001, Taiwan}

\begin{document}
\title{SALT3: An Improved Type Ia Supernova Model for Measuring Cosmic Distances}

\author[0000-0002-5153-5983]{W.~D.~Kenworthy}
\affiliation{\JHU}
\author[0000-0002-6230-0151]{D.~O.~Jones}
\altaffiliation{\Einstein}
\affiliation{\UCSC}

\author[0000-0002-5995-9692]{M.~Dai}
\affiliation{\JHU}
\affiliation{\Rutgers}

\author{R.~Kessler}
\affiliation{\Kavli}
\author{D.~Scolnic}
\affiliation{\Duke}

\author{D.~Brout}
\altaffiliation{\Einstein}
\affiliation{\Harvard}
\author{M.~R.~Siebert}
\affiliation{\UCSC}

\author{J.~D.~R.~Pierel}
\affiliation{\USC}
\altaffiliation{Now at Space Telescope Science Institute}

\author[0000-0001-7519-133X]{K.~G.~Dettman}
\affiliation{\Rutgers}

\author{G.~Dimitriadis}
\affiliation{\UCSC}

\author[0000-0002-2445-5275]{R.~J.~Foley}
\affiliation{\UCSC}

\author[0000-0001-8738-6011]{S.~W.~Jha}
\affiliation{\Rutgers}

\author[0000-0001-8415-6720]{Y.-C.~Pan}
\affiliation{\NCUG}

\author{A.~Riess}
\affiliation{\JHU}
\affiliation{\STScI}

\author{S.~Rodney}
\affiliation{\USC}

\author[0000-0002-7559-315X]{C.~Rojas-Bravo}
\affiliation{\UCSC}

\correspondingauthor{W.~D.~Kenworthy}
\email{wkenwor1@jhu.edu}
\correspondingauthor{D.~O.~Jones}
\email{david.jones@ucsc.edu}

\begin{abstract}
A  spectral-energy  distribution  (SED)  model  for  Type  Ia  supernovae  (SNe\,Ia)  is  a  critical  tool  for measuring  precise and accurate distances across a large redshift range and constraining cosmological parameters. We present an improved model framework, SALT3, which has several advantages over current models including the leading SALT2 model (SALT2.4). 
While SALT3 has a similar philosophy, it differs from SALT2 by having improved estimation of uncertainties, better separation of color and light-curve stretch, and a publicly available training code.
We present the application of our training method on a cross-calibrated compilation of 1083~SNe with 1207 spectra. Our compilation is $2.5\times$ larger than the SALT2 training sample and has greatly reduced calibration uncertainties. 
The resulting trained SALT3.K21 model has an extended wavelength range $2000$-$11000$~\AA\ (1800~\AA\ redder) and reduced uncertainties compared to SALT2, enabling accurate use of low-$z$ $I$ and $iz$ photometric bands.
Including these previously discarded bands,
SALT3.K21 reduces the Hubble scatter of the low-z Foundation and CfA3 samples by 15\% and 10\%, respectively.
To check for potential systematic uncertainties we compare distances of low ($0.01<z<0.2$) and high ($0.4<z<0.6$) redshift SNe in the training compilation, finding an insignificant $2\pm14$~mmag shift between SALT2.4 and SALT3.K21.
While the SALT3.K21 model was trained on optical data, our method can be used to build a model for rest-frame NIR samples from the \textit{Roman Space Telescope}. 
Our open-source training code, public training data, model, and documentation are available at \url{https://saltshaker.readthedocs.io/en/latest/}, and the model is integrated into the \texttt{sncosmo} and \texttt{SNANA} software packages.

\end{abstract}

\section{Introduction}
\label{sec:intro}
Type Ia supernovae (\sne) have been used as cosmological distance indicators for more than two decades, providing early evidence of the accelerating expansion of the Universe \citep{Riess1998,Perlmutter1999}. Today, \sn\ distances are used at low redshift ($z \leq 0.15 $) for distance ladder measurements of the Hubble constant \citep[H$_0$;][]{Riess2021GaiaEDR3,Freedman2019}, currently the subject of a $4-6 \sigma$ tension (see \citealp{Verde2019} for a review), as well as measurements of the dark energy equation-of-state parameter, $w$ which incorporate SNe at $z \leq 2.2 $ (currently consistent with $w = -1$;  \citealp{Scolnic2018,Abbott2019,Jones2019}). Recent measurements of $H_0$ \citep{Riess2016}, as well as most large studies of SNe across the observed redshift range for the last decade \citep{Guy2010,Conley2011,Sako2018,Betoule2014,Riess2018a, Scolnic2018,Brout2019,Jones2019}, have relied upon the SALT2 light-curve model \citep{Guy2007,Guy2010} for the brightness standardization of SNe\,Ia in their analysis. 

 \sn\ distances are typically estimated by fitting their light curves with a model to determine an overall flux, a color, and one (or more) light-curve shape parameters. The apparent magnitude (as computed from the flux) is standardized with a linear combination of color and light-curve parameters (referred to as the Tripp estimator; \citealp{Tripp1998a}) to produce a standardized apparent magnitude relative to a fiducial \sn. The SALT2 ( Spectral Adaptive Light-curve Template) model describes \sn\ light curves as a combination of component spectral energy distributions (flux surfaces defined in wavelength and time), multiplied by a color-dependent term described by a color law that is similar to that of the Milky Way. These components are determined through a ``model training'' process; the last trained model to be used in a published cosmological analysis was SALT2.4 (which we hereafter refer to as SALT2.JLA\footnote{
  JLA refers to "Joint Lightcurve Analysis" that included the SDSS-SN and SNLS teams, and produced cosmology results in \citet{Betoule2014}} ), presented in \citet{Betoule2014} and \citet{Mosher2014}, although a retrained model has recently been presented in \citet{2021arXiv210400172T}. The ubiquity of the SALT2 model in cosmology analyses of the last decade can be attributed to: 
\begin{enumerate}
    \item The spectrophotometric model can be integrated over filter bands using the appropriate rest-frame wavelengths, which removes the need for explicit $k$-corrections.
    \item The training process is cosmology independent, as the overall normalization of a light curve is a fitted parameter.
    \item The training set incorporates photometric data from multiple surveys, reducing the sensitivity of the model to the calibration of any one survey.
    \item The training sample incorporates high redshift photometric data, allowing the use of well-calibrated observer-frame optical data to extend the model into the rest-frame ultra-violet (UV).
    \item Publicly available analysis tools such as \texttt{SNANA} \citep{Kessler2009SNANA:Analysis} and \texttt{sncosmo}. \citep{Barbady2015sncosmo,Barbady2016sncosmo} include infrastructure to fit lightcurves and generate simulated data using the SALT2 model.
    \item The model has been tested by many independent groups as a consequence of its use in cosmology analyses.
\end{enumerate}

Despite these advantages, \citet{Scolnic2018} found that the calibration of the training sample used to create the SALT2.JLA model  was the largest single systematic uncertainty in their measurement of $w$ ($\sigma_w=0.014$, 30\% of the total systematic uncertainty), though a new analysis methodology could somewhat reduce this uncertainty \citep{Brout2020binning}. Achieving the science goals of the Vera C.\ Rubin Observatory's Legacy Survey of Space and Time (LSST) will require that systematic uncertainty in the calibration of the light-curve model be decreased by a factor of 5 for the year-one analysis \citep{TheLSSTDarkEnergyScienceCollaboration2018}.

The SALT2.JLA model does not fully reproduce observed spectral features such as varying absorption line velocities \citep{Foley2011MeasuringDistances,Siebert2020}.  Similarly, studies have found evidence that standardized distances  based on SALT2 with the Tripp estimator (SALT2+Tripp) are dependent on \sn\ host galaxy properties \citep{Kelly2010,Sullivan2010}, although the best way to characterize this effect is still in question \citep{Rigault2013,Betoule2014,Jones2018,Rigault2018,Brout2020,Smith2020HostGalaxies}. Since these astrophysical effects are not explicitly included in the training process, differences between the training sample and cosmology samples can lead to subtle biases in the cosmology results.  To characterize and correct for such untrained effects, it is  essential to perform training studies on simulated data that incorporates a broad range of physical effects. Therefore, in parallel with the the SALT3 development our team has developed a more general SED-simulation framework described in  \citet{Pierel2020}. 

Extending the wavelength range of the SALT model shows promise in the reduction of statistical uncertainty. Distance standardization based on  SALT2+Tripp results in scatter about the Hubble diagram that is $\sim \SI{0.1}{\mag}$ larger than expected from photometric uncertainties (which we refer to as ``intrinsic scatter''). Numerous studies over the last two decades have found that NIR peak magnitudes show smaller Hubble residuals than SALT2+Tripp distances \citep{Krisciunas2004,Krisciunas2007,WoodVasey2008,Burns2011,Mandel2011,Dhawan2018,Avelino2019,Mandel2020}. However, the SALT2.JLA SED model extends only to $9200$~\AA\ (extrapolated further for use in simulations in \citealp{Pierel2018Extending}) with substantial model uncertainties past $\sim 7000$~\AA\ that preclude the use of existing low-redshift optical data. By extending the wavelength range to reliably include existing rest-frame $i$ and $z$ band photometry we can improve on cosmology constraints on current data sets. Further extension of the model into the NIR would allow future SN Ia cosmology programs to make use of a wavelength  range in which SNe\,Ia are intrinsically more precise.

 As a step toward these goals, we have defined a SALT3 spectrophotometric model formalism and developed \texttt{SALTshaker}, a flexible and open source Python-based code for training a SALT3 model, accepting both spectra and photometry in the training process. The SALT3 formalism has been defined similarly to SALT2 to retain the compatibility of our model with existing analysis frameworks. As part of an overall testing and validation framework (Dai et al.\ in prep), \texttt{SALTshaker} enables new SN Ia light-curve models to be quickly trained for new samples of cosmological SNe, allowing uncertainties from the modeling process to improve as larger and more accurately calibrated samples are collected.  In Section \ref{sec:model} we define our SALT3 model, and describe the procedure of our training code. In Section \ref{sec:valid} we apply \texttt{SALTshaker} on training data similar to that of the SALT2.JLA model, allowing a direct comparison of our training process to that of SALT2. Next we add recalibrated data from past and recent \sn\ surveys to our training sample, described in Section \ref{sec:data}, to increase the size of the training sample by a factor of $\sim 2.5$. Finally, we build the SALT3.K21 model and present our results in Section \ref{sec:extended_salt}.

\section{The SALT3 Model and \texttt{SALTshaker}}
\label{sec:model}

\begin{figure}
    \centering
    \includegraphics[width=3.5in]{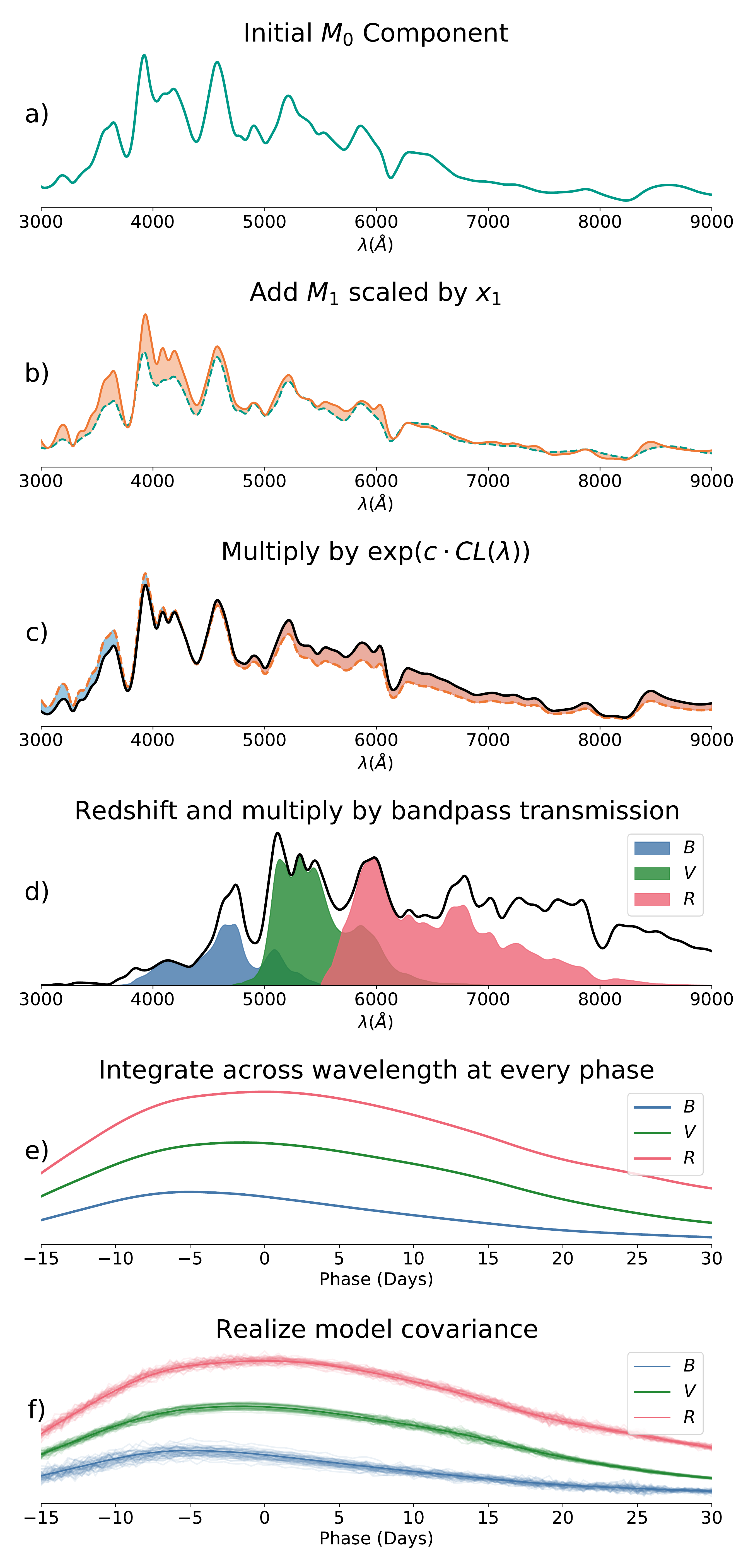}
    \caption{Construction of a SALT light curve for a given \sn, here shown using the SALT2.JLA model. In panels a-c, we apply the $M_1$ component and the color law to the base $M_0$ spectral flux at a phase of +10 days. In panel d, the spectrum is redshifted, and integrated across a bandpass transmission. In panel e, this process is repeated at every phase to produce a lightcurve. Panel f) shows realizations of the model uncertainties. Color scatter can be seen as the coherent offset within each band, clearly seen at the peak of $V$ band. At later phases in $V$ the uncorrelated uncertainties dominate the realizations. }
    \label{fig:saltlightcurvecartoon}
\end{figure}

Using the framework of SALT, the SALT3 model defines the spectral flux of a \sn\ as a function of rest-frame wavelength $\lambda$ and phase $p$. Given three light-curve parameters for a given SN $x_0$ (the overall flux normalization), $x_1$ (associated with the light-curve ``stretch'' ), and $c$ (a fitted color parameter), the spectral flux\footnote{In all cases we use $F$ to refer to a spectral flux and $f$ to refer to a flux integrated over an appropriate bandpass} (in energy units) is
\begin{align} \label{eq:spectralfluxmodel}
    F(p,\lambda) = &x_0 [M_0(p,\lambda;\boldsymbol{m_0}) + x_1 M_1(p,\lambda;\boldsymbol{m_1})] \nonumber\\ & \cdot \exp(c \cdot CL(\lambda;\boldsymbol{cl})),
\end{align}
\noindent where $M_0(p,\lambda;\boldsymbol{m_0})$ and $M_1(p,\lambda;\boldsymbol{m_1})$ are flux surfaces similar to principal components, representing respectively the SED of a fiducial \sn\ and a first order correction. The two flux surfaces are defined on a basis of second-order B-splines, with $N_\text{Phase} \times N_\lambda$ basis functions for each surface, with coefficients $\boldsymbol{m_0},\boldsymbol{m_1}$ that are model parameters. The number of basis functions along each axis are determined based on the desired wavelength and phase resolution; in our work we have chosen wavelength resolution $70$~\AA\ and phase resolution of $3$ days. $CL(\lambda;\boldsymbol{cl})$ is a single color law which combines the effects of intrinsic color variation and host galaxy dust extinction. The color law is polynomial with $N_{CL}$ coefficients $\boldsymbol{cl}$ within a specified  wavelength range $\lambda_{-}$ to $\lambda_{+}$ and linearly extrapolated for the rest of the SED wavelength range of the model. Thus the color law is defined as
\begin{align*}
  CL_p(\lambda) &= \sum_{i=0}^{N_{CL}} \boldsymbol{cl}_i \lambda^i, &&  \\
  & && \lambda <\lambda_{-} \\
  CL(\lambda) &= 
    \smash{\left\{\begin{array}{@{}l@{}}
      CL'_p(\lambda_{-}) (\lambda -\lambda_{-}) +CL_p(\lambda_{-}) \\[\jot]
      CL'_p(\lambda_{+}) (\lambda -\lambda_{+}) +CL_p(\lambda_{+}) \\[\jot]
      CL_p(\lambda)
    \end{array}\right.} && \lambda >\lambda_{+} \\
  & && \text{otherwise}
  \label{eq:colorlaw}
\end{align*}
\noindent  The object of the model training process is to determine the model parameters $\{\boldsymbol{m_0},\boldsymbol{m_1},\boldsymbol{cl}\}$ and estimate un-modeled variability in the flux surfaces; the model training is detailed in Section \ref{subsec:trainingprocedure}. 

 For a given photometric observation of a \sn\ at heliocentric redshift $z_\text{Hel}$ in a filter $X$ with observer-frame transmission function $T_X(\lambda)$ in photon units, the broadband flux as a function of phase is 
\begin{align}
    f_X(p)=\int T_X(\lambda/(1+z_\text{Hel})) F(p,\lambda; x_0,x_1,c) \lambda d \lambda
\end{align}
\noindent  We illustrate this procedure for modeling a light curve in Figure \ref{fig:saltlightcurvecartoon}. The fixed configuration parameters used to specify the model are listed in Table \ref{table:modelconfig}, while the fitted model parameters ( $\{\mathbf{m_0},\mathbf{m_1},\mathbf{cl}\}$) are listed in Table \ref{table:allparameters}.

\subsection{Model Definitions}
\label{subsec:definitions}
The specification of the model given above is degenerate; for example, the scale of the flux surfaces may be changed by reducing $M_0$ and $M_1$ and increasing $x_0$ by the same factor. To remove degeneracies we apply further model definitions, which are used as constraints during the training process. These definitions are arbitrary, and have been chosen to define a ``fiducial'' supernova whose SED is the $M_0$ flux surface at the mean of the observed distributions of lightcurve parameters. The definitions are:
\begin{enumerate}
    \item The rest-frame synthetic $B$-band flux of the $M_0$ component at peak is fixed such that $m_B^\text{peak}=10.5$ when $x_0=1$
    \item The rest-frame synthetic $B$-band flux of the $M_1$ component at peak is defined to be 0
    \item The distribution of the light-curve parameter $x_1$ in the training sample is defined to have 0 mean  
    \item The distribution of $x_1$ has standard deviation 1
    \item The distribution of the light-curve parameter $c$ in the training sample is defined to have 0 mean 
    \item The color law is defined such that $CL(4300 \textrm{ \AA}) = 0$ and $CL(5430 \textrm{ \AA})= -1$ , corresponding to central wavelengths for $B$ and $V$ bandpasses
    \item The distributions of $x_1$ and $c$ have no correlation in the training sample
\end{enumerate}

The location and scale of the $x_1$ distribution inferred for any cosmological sample are thus \textit{relative} to the demographics of the training sample for a particular model, while $c$ is fixed to correspond to a phase-independent inferred $B-V$ color relative to the mean of the training sample. Our model definitions and SALT2 differ only in the last definition, intended to separate the phase-independent color from the phase-dependent $M_1$ component. The SALT2 training code does not constrain the correlation between $x_1$ and $c$, and instead fixes the $V$-band flux (and $B$-band flux following definition 2) of the $M_1$ component to be $0$ at peak brightness, implying that SALT2 $x_1$ has no effect on observed $B-V$ color at peak. For SALT3, removing the correlation between the parameters $x_1$ and $c$ has a physically intuitive meaning; the dust-like color term does not depend on the processes associated with light-curve stretch. Further, it is easier to make inferences about the latent populations when the stretch and color parameters are uncorrelated, as correlated parameters imply that there is redundant information in the two parameters.

\subsection{Photometric model uncertainties}

As \citet{Rubin2020Dimensionality} and \citet{Rose2020SNEMOEval} suggest that \sn\ SEDs are determined by 3-5 SED parameters and a color, a model with a single SED parameter ($x_1$ in our case) will not capture the full diversity of the population. We refer to the variation unexplained by our model as ``in-sample variance''. This variation can be contrasted with the uncertainty in the model parameters due to training with a finite sample of SNe with finite signal-to-noise photometry, which we refer to as ``out-of-sample variance''. For a photometric observation in a filter with central wavelength $\lambda_c$ and phase $p$, in-sample variance is addressed by a ``model variance'' composed of two terms. The first is a diagonal uncertainty defined as 
\begin{align}
    \sigma^2_f(p,\lambda_c)=& \left[ x_0 \exp(c \cdot CL(\lambda_c))  \int T \left(\frac {\lambda}{1+z_\text{Hel}}\right) \lambda d\lambda \right]^2 \nonumber \\& \times \left[\sigma^2_{M_0}(p,\lambda_c )+ x^2_1 \sigma^2_{M_1}(p,\lambda_c ) +\right. \nonumber \\
    & \left. 2 x_1 C_{M_0 M_1} (p,\lambda_c) \sigma_{M_0}(p,\lambda_c )  \sigma_{M_1}(p,\lambda_c )  \right]
\end{align}
where $\sigma_{M_0}(p,\lambda_c ;\bold{\sigma_{M_0}})$ and $\sigma_{M_1}(p,\lambda_c ;\bold{\sigma_{M_1}})$ represent variability in the flux surfaces and $C_{M_0 M_1} (p,\lambda_c;\bold{C_{M_0,M_1}})$ is the correlation between flux surfaces. These variance terms are described by zeroth order B-splines (equivalent to binning the data by phase and wavelength) with 8 basis functions in wavelength and 12 basis functions in phase. As detailed in Section \ref{subsec:trainingprocedure} we use a maximum likelihood estimator to determine the B-spline coefficients $\boldsymbol{\sigma_{M_0}}, \boldsymbol{\sigma_{M_1}},\boldsymbol{C_{M_0,M_1}}$. This is distinct from the approach of SALT2, which took the in-sample variance to have the same form as that of the out-of-sample variance, scaling the latter (evaluated by leave-one-out tests) using a smooth function or ``error snake'' to fix the $\chi^2_\nu$ of the photometry of the training sample to 1. Because the in-sample variance is determined by variability of the underlying population of SNe light curves while the out-of-sample variance is determined by the distribution of available training data, we consider our approach to be a better account of the remaining variance of the SN Ia population.

The second term of the in-sample variance is a covariant ``color scatter'' that allows light curves of the same supernova in different bands to be coherently offset relative to one another. The color scatter term between photometric measurements using the same bandpass is correlated, with no correlation between measurements in different bandpasses. This model is similar to chromatic models of intrinsic scatter like those of \citet{Guy2010} and the diagonal terms of the covariance matrix from \citet{Chotard2011}. The  relative covariance $k^2(\lambda_c;\mathbf{a})$ is described by the exponential of a fourth order polynomial in wavelength, where the polynomial coefficients $\mathbf{a}$ are fitted parameters. Thus, the model covariance matrix for two photometric observations $i$ and $j$ in broadband filters $X_i$,$X_j$ of a given supernova with model fluxes $\Vec{f}_\text{Model}$ is
\begin{align}
    (\Sigma_{\text{Model}}) _{ij} = &\delta_{ij} \sigma^2_f(p_i,\lambda_{c (i)}) \nonumber \\+  &\begin{cases} 
        k^2([\lambda_c]) (\Vec{f}_\text{Model})_i (\Vec{f}_\text{Model})_j & X_i = X_j\\
        0 & \text{otherwise}
    \label{eq:photcovariance}
\end{cases}
\end{align}
\subsection{Modeling of Spectral Data}
Although the SALT3 model is intended for use as a photometric light-curve model, spectral data is included in the training process to better constrain the shape of spectral features. However, given that spectral data have larger calibration uncertainties compared to broadband fluxes, we follow the SALT2 training code  in ``recalibrating'' spectral data, modulating the model by a smooth function to match the continuum of the observed spectrum. We modify the spectral flux equation (Equation \ref{eq:spectralfluxmodel}) for use with spectra during the training by removing the color term and replacing it with a recalibration term of similar form,

\begin{equation}
    F_\text{spec} (p,\lambda) = x_0 [M_0(p,\lambda) + x_1 M_1(p,\lambda)] \exp( \sum^n_i y_i \lambda^i /i!),
\end{equation}

\noindent where the spectral recalibration nuisance parameters $y_i$ are fitted during the training procedure. In the limit of perfectly calibrated spectra, the fitted recalibration term will reproduce the effect of the color parameter on the spectrum; by removing the color law entirely from the spectral flux equation, we mitigate the impact of mis-calibrated spectra on the color law. The quantity $n$ controls how many recalibration parameters are allowed for each spectrum, and is determined based on the wavelength extent of the spectrum and the number of filter bands available for that SN (additional filter bands better constrain the recalibration term, allowing for more free parameters). The model variances are defined similarly to the photometry, without the contribution of the color scatter:

\begin{align}
    \sigma^2_F(p,\lambda)=& \left[ x_0 \exp( \sum^n_i y_i \lambda^i /i!))\right]^2 \times \nonumber \\& [\sigma^2_{M_0}(p,\lambda )+ x^2_1 \sigma^2_{M_1}(p,\lambda ) \nonumber \\
    +&2 x_1 C_{M_0 M_1} (p,\lambda) \sigma_{M_0}(p,\lambda )  \sigma_{M_1}(p,\lambda )  ].
    \label{eq:specuncertainty}
\end{align}

\begin{table*}
    \centering
    \caption{Fixed Model Parameters}
    \begin{tabular}{lrrr}
        \hline \hline\\[-1.5ex]
        Parameter & Description & JLA Training & SALT3 Training\\
        \hline\\[-1.5ex]
        $A_\text{Spec}$ & Spectral Suppression & \nodata & 0.75\\
        $N_\text{Phase}$ & Number of phase basis functions for flux surfaces & 11 & 20\\
        $N_\lambda$ & Number of basis functions on wavelength axis for flux surfaces & 97 & 127\\
        $N_{\sigma,\text{Phase}}$ & Number of phase basis functions for flux uncertainty surfaces & \nodata & 9\\
        $N_{\sigma,\lambda}$ & Number of basis functions on wavelength axis for flux uncertainty surfaces & \nodata & 5\\
        $A_\text{Phase}$ & Phase gradient regularization weight & 0 & 1000\\
        $A_\lambda$ & Wave gradient regularization weight & 10 & 10000\\
        $A_\text{Dyadic}$ & Dyadic regularization weight & 1000 & 1000\\
        \hline\\[-1.5ex]
    \multicolumn{4}{l}{
    \begin{minipage}{7in}
    Model parameters for SALT2.JLA training and SALT3.K21 training.  We note that given our adjusted regularization scheme, regularization weights are not directly comparable.  We were unable to determine the exact spectral suppression in the JLA training, but from \citet{Mosher2014} we expect that this is fine-tuned for a given training sample.
    \end{minipage}}
        \end{tabular}
    \label{table:modelconfig}
\end{table*}

\begin{figure*}
    \centering
    \includegraphics[width=7in]{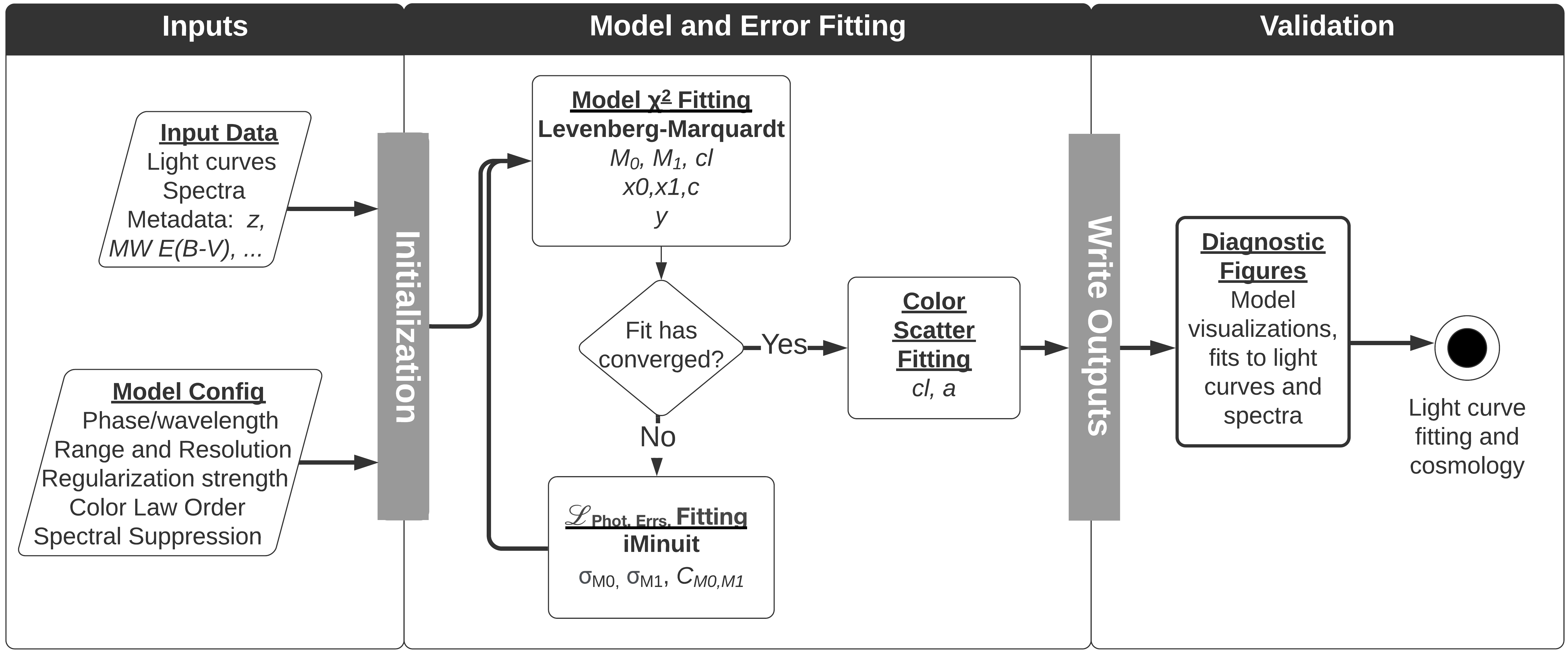}
    \caption{Activity diagram of the \texttt{SALTshaker} training process, with variable names in the model and model error fitting sections corresponding to the descriptions in Table \ref{table:allparameters}.}
    \label{fig:saltshaker_schematic}
\end{figure*}

\begin{table*}[]
    \centering
    \caption{Model and Model Uncertainty Parameters}
    \begin{tabular}{lrrr}
        \hline \hline\\[-1.5ex]
        Parameter & Category & Number & Description\\
        \hline\\[-1.5ex]
        $\bold{m_0}$ & Flux Model & $N_\text{Phase} \times N_\lambda$ & $M_0$ B-spline coefficients \\
        $\bold{m_1}$ & Flux Model & $N_\text{Phase} \times N_\lambda$ & $M_1$ B-spline coefficients \\
        $\bold{cl}$ & Flux Model & $N_{CL}$ & Color law \\
        $x_0$ & Nuisance & $N_\text{SN}$ & Overall flux normalization for each SN \\
        $x_1$ & Nuisance & $N_\text{SN}$ & Stretch of each SN \\
        $c$ & Nuisance & $N_\text{SN}$ & Color for each SN \\
        $\bold{y}$ & Nuisance & $\propto N_\text{spec} $ &  Spectral recalibration\\
        $\bold{\sigma_{M_0}}$ & Uncertainty & 72 & Uncertainty in $M_0$ \\
        $\bold{\sigma_{M_1}}$ & Uncertainty & 72 & Uncertainty in $M_1$ \\
        $\bold{C_{M_0,M_1}}$ & Uncertainty & 72 & Correlation between $M_0$ and $M_1$ \\
        $\bold{a}$ & Color Scatter & 4 & Color scatter \\
        \hline\\[-1.5ex]
        \hline\\[-1.5ex]
    \end{tabular}
    \label{table:allparameters}
\end{table*}

\subsection{ Regularization}
In regions of phase-wavelength space that are poorly constrained by spectra, the $M_0$ and $M_1$ components can acquire artifacts such as high-frequency ringing, or deconvolution noise. To reduce these artifacts, we use a ``regularization'' procedure that penalizes large derivatives in the model where there is an absence of spectroscopic data, as parameterized by a binned spectral density function $\neff(p,\lambda)$, with every spectrum assigned equal weight. We implement three kinds of regularization: phase gradient, wave gradient, and dyadic. These are applied to the two flux surfaces $M_0(p,\lambda)$ and $M_1(p,\lambda)$. For a flux surface $S(p,\lambda)$ the regularization terms are defined as

\begin{enumerate}
    \item Phase gradient regularization penalizes large derivatives with respect to phase in less-constrained regions
    \begin{align}
        \chi^2_\text{Phase}&[S(p,\lambda) ] = A_\text{Phase}\sum^{2N_\text{Phase}}_i \sum^{2N_\lambda}_j \frac{{\Big (}\left.\frac{\partial S}{\partial p}\right|_{\substack{p=p_i\\ \lambda=\lambda_j}}{\Big )}^2}{{\neff}(p_i,\lambda_j)} 
    \end{align}
    \item Wave gradient regularization similarly penalizes large derivatives with respect to wavelength in less-constrained regions
    \begin{align}
        \chi^2_\text{Phase}&[S(p,\lambda) ] = A_\text{Wave}\sum^{2N_\text{Phase}}_i \sum^{2N_\lambda}_j \frac{{\Big (}\left.\frac{\partial S}{\partial \lambda}\right|_{\substack{p=p_i\\ \lambda=\lambda_j}}{\Big )}^2}{{\neff}(p_i,\lambda_j)} 
    \end{align}
    \item Dyadic regularization encourages the flux surfaces to be separable in phase and wavelength, and is 0 when a flux surface takes the form $S(p,\lambda)=g(p) \times h(\lambda)$
    \begin{align}
        \chi^2&_\text{Dyadic}[S(p,\lambda) ] = A_\text{Dyadic} \sum^{2N_\text{Phase}}_i \sum^{2N_\lambda}_j \frac{1 }{{\neff}(p_i,\lambda_j)}\nonumber \\&  \times {\Big [} \left.\frac{\partial S}{\partial p}\right|_{\substack{p=p_i\\ \lambda=\lambda_j}} \left.\frac{\partial S}{\partial \lambda}\right|_{\substack{p=p_i\\ \lambda=\lambda_j}} -  S(p_i,\lambda_j)  \left.\frac{\partial^2 S}{\partial p \partial \lambda }\right|_{\substack{p=p_i \\ \lambda=\lambda_j}} {\Big ]}^2
    \end{align}
\end{enumerate}

The relative strength of the three regularization terms is determined by the weights $\{A_\text{Phase}, A_\text{Wave}, A_\text{Dyadic}\}$, which are inputs to the \texttt{SALTshaker} code.  The summations over phase and wavelength points are evenly spaced over the model phase and wavelength ranges, with twice as many points as basis functions along each axis. As regularization can bias the model surfaces by over-smoothing them \citep{Mosher2014}, we tune the weights to ensure that the regularization terms do not contribute significantly to the total $\chi^2$. Further work to determine how this regularization scheme affects the model and to choose optimal model configurations will require applying our training and analysis framework to simulations (Dai et al.\ in prep; \citealp{Pierel2020}).

\subsection{Training Procedure}
\label{subsec:trainingprocedure}

Based on the model definitions above, we construct the training $\chi^2$ for the model as
\begin{align}
    \chi^2_\text{Total}=&\chi^2_\text{Phot}+A_\text{Spec} \chi^2_\text{Spec}+\chi^2_\text{Constraints}+\chi^2_\text{Reg}. 
\end{align} 
The photometric $\chi^2$ term is
\begin{align}
    \chi^2_\text{Phot}=& \sum^{N_\text{SN}}_{n} (\Vec{f}_\text{obs}^{(n)} - \Vec{f}_\text{Model}^{(n)})^T(\Sigma_\text{Total}^{(n)}) ^ {-1} (\Vec{f}^{(n)}_\text{obs} - \Vec{f}^{(n)}_\text{Model})  \nonumber\\
    \Sigma_\text{Total}^{(n)} &=\text{diag}((\sigma^{(n)}_{Phot})^{2}) + \Sigma_{\text{Model}}^{(n)}
\end{align}
where for the $n$th supernova $\Vec{f}^{(n)}_\text{obs}$ is the vector of observed photometric fluxes, and $\Vec{f}^{(n)}_\text{Model}$ is the vector of the model fluxes integrated over the photometric bandpass.  The covariance $\Sigma_\text{Total}^{(n)}$ combines  photometric uncertainties and the model uncertainty covariance described in Equation \ref{eq:photcovariance}. The factor $A_\text{Spec}$ is a constant ``spectral suppression'' term that downweights the contribution of spectra to the training $\chi^2$ to reduce the sensitivity of the training to unknown systematic errors in the spectral data \citep{Guy2007}. We set this term such that the spectral and photometric data have roughly equal contributions to the total training $\chi^2$. The spectral $\chi^2$ term is then defined
\begin{align}
    \chi^2_\text{Spec}=\sum^{N_{SN}}_n \sum^{N_\text{spec}}_i \sum^{N_\text{points}}_j \frac{[F_\text{spec} (p^{(n,i)},\lambda^{(n,i)}_j) - (\Vec{f}^{(n,i)}_\text{obs})_j]^2}{(\sigma_\text{obs})_j^2+\sigma_\text{Model}(p^{(n,i)},\lambda^{(n,i)}_j)^2},
\end{align}

\noindent where for the $n$th supernova, $F_\text{spec} (p^{(n,i)},\lambda^{(n,i)}_j)$ is the model spectral flux at the $j$th wavelength bin of the $i$th spectrum,  $(\Vec{f}^{(n,i)}_\text{obs})_j$ is the observed spectral flux, $(\sigma_\text{obs})_j$ is the photometric uncertainty, and $\sigma_\text{Model}(p^{(n,i)},\lambda^{(n,i)}_j)$ is the model uncertainty in Equation \ref{eq:specuncertainty}.  In $\chi^2_\textrm{phot}$ and $\chi^2_\textrm{spec}$, we redden the model to account for the Milky Way $E(B-V)$ along the line of sight to each SN, but do not include this term in the equations above for simplicity. The $\chi^2_\text{Constraints}$ is composed of penalty terms used to enforce the model definitions described in Section \ref{subsec:definitions}, and the regularization term is defined as 
\begin{align}
    \chi^2_\text{Reg}=\sum^{\{M_0,M_1 \}}_{S(p,\lambda) } & \chi^2_\text{Phase}[S(p,\lambda)]+\chi^2_\text{Wave}[S(p,\lambda) ] \nonumber\\
    &+ \chi^2_\text{Dyadic}[S(p,\lambda) ] .
\end{align}

The \texttt{SALTshaker} code is initialized with the configuration parameters shown in Table \ref{table:modelconfig}. \texttt{SALTshaker} then determines best-fit values of the model parameters. These parameters are shown in Table \ref{table:allparameters}. We define flux model parameters as those which control the flux surfaces and color law, uncertainty model parameters as those which control the output model uncertainties, and nuisance parameters as those which describe individual supernovae. 

 \texttt{SALTshaker} alternates between simultaneously fitting the nuisance parameters and model flux parameters while keeping uncertainties fixed and fitting the model uncertainty parameters while keeping model fluxes fixed.  The color scatter is kept fixed at $k(\lambda_c)=0$ during this stage of the training process, as we find that allowing nonzero color scatter results in biased flux surfaces due to the regularization procedure. The flux and nuisance parameters are fit using an iterative Levenberg-Marquardt algorithm. To reduce the number of (computationally expensive)  Jacobian evaluations of the model residuals required, we use Schubert's method to perform a rank one update on the Jacobian after each iteration while maintaining its sparsity structure \citep{Marwil1979,Schubert1970}. However this technique is unsuitable for determining the model uncertainties.

To fit the model uncertainty parameters  defined in Equation \ref{eq:photcovariance} we define a  log-likelihood $(\mathcal{L}_\text{Phot. Errs.}$ given as
\begin{align}
   - 2\log(\mathcal{L}_\text{Phot. Errs.})= & \chi^2_\text{Phot} - \sum^{N_{SN}}_n \log(\lvert \Sigma^{(n)}_\text{Total} \rvert).
\end{align}
The model uncertainty parameters are chosen to maximize the log-likelihood using the optimizer iMinuit while the flux model and nuisance parameters are kept fixed \citep{James1975,dembinski2020}. After the training has converged with $k$ fixed to 0, we use iMinuit to fit the parameters controlling the color scatter; during this final fit, the color law is allowed to vary (having previously been fit as a flux model parameter). The out-of-sample variance is then estimated by inverting the Hessian matrix of the flux parameters obtained from the model fitting process, with the regularization terms suppressed by a factor of 100, and propagating those parameter uncertainties into the flux surfaces. The model uncertainties and out-of-sample variance are added together as the total model uncertainty surface.

The computation time required for the SALT3.JLA training sample with \texttt{SALTshaker} is approximately three minutes per iteration, with $\sim$25 iterations required for convergence.  On the larger SALT3.K21 sample, iterations are significantly longer, with $\sim$25 minutes required per iteration. We note that 29-\AA\ spectral binning, which was used in the SALT2.JLA training, improves speeds substantially by reducing the amount of data by up to an order of magnitude.  We include this as an option in \texttt{SALTshaker} but the models in this work use the native binning of the input spectra.  Finally, the slowest component of the code is the iterative fitting of the error model, which for SALT3.K21 requires approximately 4 hours to perform 80 iterations and reach convergence through iMinuit.  Error model iterations are performed every five iterations by default but could likely be performed less often without adversely affecting the final model.  Faster methods of error model fitting will be an important avenue for future improvement and should improve speeds considerably.

An overview of the \texttt{SALTshaker} training procedure is given in Figure \ref{fig:saltshaker_schematic}.

\section{\texttt{SALTshaker} Validation}
\label{sec:valid}

Here, we show that our method is capable of producing a trained model that reproduces inferred distances from SALT2 by comparing trained SALT3 models with SALT2.JLA. We show comparisons between synthetic photometry in multiple rest-frame filters as well as the spectral models, and compare distances between the models.
We describe our metrics in Section \ref{subsec:validationprocedure}, the JLA training sample of \citet{Betoule2014} and \citet{Mosher2014} in \ref{subsec:jlasaple}, a simulated sample that mimics the demographics of JLA in \ref{subsec:jlasim}, then train on these samples in \ref{subsec:simjlatraining}, \ref{subsec:jla_orig_training}.


\subsection{Validation Procedure and Metrics}
\label{subsec:validationprocedure}
To distinguish between models trained on different input data, we define ``SALT3.\textit{X}'' as the SALT3 model created with \texttt{SALTshaker} using training sample \textit{X}. We refer to the samples used to evaluate the performance of a given SALT3 model as ``validation'' samples, as it will be expedient to evaluate the trained model in some cases on data that was not used in the model training. We use the \texttt{SNANA} light curve fitting program to fit validation samples with both SALT2.JLA and SALT3.\textit{X}. 

Given fitted SALT parameters $m_B,x_1,c$ and their uncertainties, the Tripp estimator for distance modulus is
\begin{equation}
    \mu = m_B + \alpha \cdot x_1 - \beta \cdot c -\mathcal{M}
\end{equation}
\noindent with distance uncertainties
\begin{align}
    \sigma_\mu=& \sigma_\textrm{int}^2+ \sigma_{\mu,z}^2 +\sigma_\textrm{lens}^2 +\sigma_{m_B}^2+ (\alpha  \sigma_{x_1})^2 + (\beta \sigma_c)^2\nonumber \\\
    &+ 2 \alpha \beta \sigma_{c,x_1}+2 \alpha \sigma_{m_B,x_1}+2\beta \sigma_{m_B,_c}
\end{align}
\noindent where $\sigma_\textrm{int},\alpha,\beta$ and $\mathcal{M}$ are nuisance parameters, $\sigma_{\mu,z}$ is computed from a peculiar velocity uncertainty of  \SI{250}{\kilo\meter\per\second}, and $\sigma_\textrm{lens}=0.055 z$. We use the {\tt SALT2mu} method \citep{Marriner2011,Kessler2017} implemented in \texttt{SNANA} to estimate the nuisance parameters as well as distances in 5 redshift bins. Allowing for a shift in location and scale of the light-curve parameters, the observed distributions are similar as compared to SALT2.JLA (see Section  \ref{subsec:lcparcomparison}).  We thus expect selection biases to be common between the two models, and therefore we do not use \texttt{SALT2mu} to correct for selection biases. 

Given estimated distance moduli $\mu$ and Hubble residuals relative to a nominal $\Lambda$CDM cosmology ($\Delta \mu$) for both models we define two metrics. First $\textrm{RMS}(\Delta \mu)$ across the validation sample and the relative distance difference between models defined as 
\begin{align}
    \textrm{Diff} (\Delta_z \mu) &=\Big[ \mu(0.40 <z <0.6 |\textrm{SALT3.\textit{X}})\nonumber \\ 
    &-  \mu(0.40 <z <0.6| \textrm{SALT2.JLA}) \Big] \nonumber\\
    &-  \Big[\mu(0.01<z<0.2 |\textrm{SALT3.\textit{X}}) \nonumber \\ 
    & - \mu(0.01<z<0.2 |\textrm{SALT2.JLA})  \Big],
\label{eq:distancediff}
\end{align}
 where $\mu( Z | M)$ indicates a weighted average distance across a redshift range $Z$ given a model $M$. Additionally we will show binned Hubble residuals across the redshift range. We discuss differences in the nuisance parameter $\beta$, as this has physical implications for dust properties, however $\alpha$ depends on both the demographics of the training sample and our revised separation of stretch and color, and thus has no useful comparison across models (see Section \ref{subsec:definitions}). Similarly we compare synthetic photometry, but these comparisons are most relevant for simulated data when the truth model is known. We train on multiple simulated and real data samples to assess how our models perform; in Table \ref{table:abbreviations} we summarize the abbreviations used for these training samples. 

\begin{table*}[tbp]
    \centering
    \caption{Data Abbreviations}
    \begin{tabular}{lrp{3in}rr}
        \hline \hline\\[-1.5ex]
        Abbreviation & Size & Description &Data&Trained Model\\
        \hline\\[-1.5ex]
        JLA & 420 & Mixture of low-redshift SNe from many surveys and high-redshift SNe from SNLS and SDSS; see \citet{Mosher2014}  &\ref{subsec:jlasaple}& \ref{subsec:jla_orig_training}\\
        simJLA & 420 & Simulated data that emulates the JLA training sample &\ref{subsec:jlasim}& \ref{subsec:simjlatraining} \\
        K21 & 1083 & Compilation of the JLA training sample combined with new SNe from Foundation, Pan-STARRS, the Dark Energy Survey, CfA4, and CSP, with additional spectra from \texttt{kaepora} & \ref{sec:data}&\ref{subsec:completesampletraining}\\
        K21train & 541 & Half of the SNe from the K21 compilation chosen at random for use as a training sample & \ref{sec:data}&\ref{subsec:separatetrainvalid} \\
        K21valid & 541 & The other half of the K21 SNe, used as a validation sample to see how the SALT3 model performs on data that was not part of the training & \ref{sec:data}&\nodata\\
\hline\\[-1.5ex]
    \end{tabular}
    \begin{minipage}{6in}
    Abbreviations we use to refer to compilations of cosmological \sne\ in this work, number of SNe in each, a brief description, the section in which we discuss the data itself, and the section in which we discuss a SALT3 model trained on each
    \end{minipage}
    \label{table:abbreviations}
\end{table*}

\begin{figure*}[btp]
    \centering
    \includegraphics[width=0.9\textwidth]{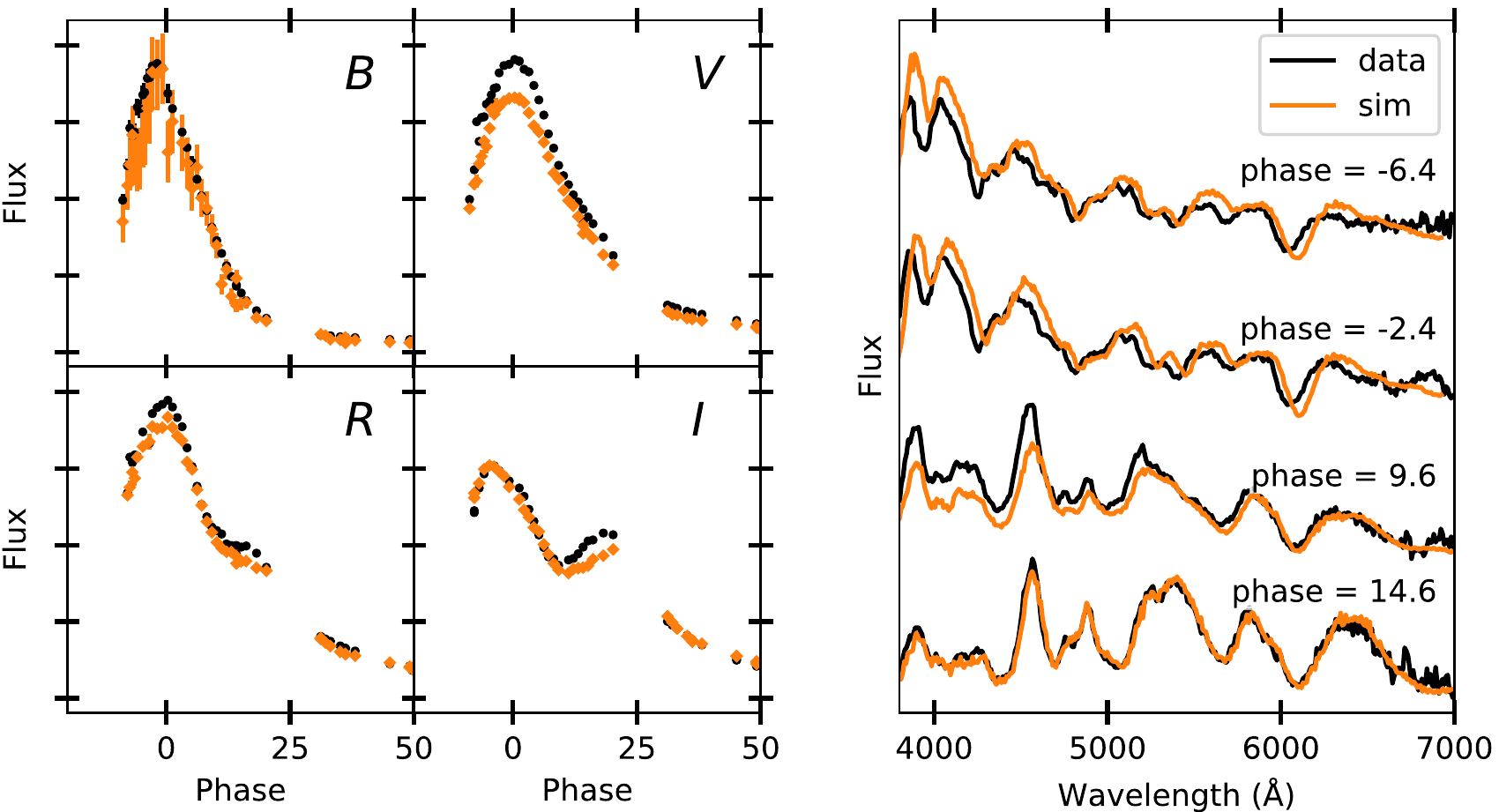}
    \caption{SN 1992A light curves (left) and spectra (right) for real data (black) and \texttt{SNANA} simulation (orange) generated with best-fit  $x_1$, $c$, $z$, and time of maximum light.  Photometric data are shown on the left and spectroscopic data on the right.  Random offsets between data and simulations are expected due to the C11 scatter model and spectral recalibration.  The simulated spectra at red wavelengths have better SNR than the data because we model the average spectral SNR rather than its wavelength dependence.}
    \label{fig:jlasim}
\end{figure*}

\subsection{JLA Training Sample}
\label{subsec:jlasaple}
The JLA data used to train the SALT2.JLA model consist of 420 SNe with light curves, 83 of which include spectroscopy, from a compilation of low-$z$ samples (see Table \ref{table:trainingsamp} for details and citations), the Sloan Digital Sky Survey (SDSS; \citealp{Holtzman08,Kessler2009,Sako2018}), and the Supernova Legacy Survey (\citealp{Astier2006}, with spectra from \citealp{Walker2011,Balland2018} and private communication with M.\ Betoule, C.\ Balland).  These data are summarized in Table \ref{table:trainingsamp}, along with the new training data discussed in Section \ref{sec:data}, and are included in the data release at \url{http://saltshaker.readthedocs.io/}.

We use the ``Supercal'' procedure  \citep{Scolnic2015} to update the calibration of the JLA training sample. Supercal uses the 3$\pi$ sky coverage of the PS1 system and its observations of secondary standard stars to precisely determine the offsets between different photometric systems. \citet{Scolnic2015} found that applying these calibration corrections to the sample of \sn\ used to measure the Hubble flow {\it without} updating the training calibration could shift $w$ by 0.026. While we do not study the impact of the calibration on the trained model, \citet{2021arXiv210400172T} examines this with  a full reanalysis of the DES-SNIa cosmology results using a retrained SALT2 model.  We also use the \citet{Schlafly2011} corrections to the \citet{Schlegel1998} Milky Way dust maps.

\subsection{Simulated JLA Training Sample}
\label{subsec:jlasim}

 

To test \texttt{SALTshaker} with a known input model and cosmology, we first {\it simulate} the SALT2.JLA training data to produce our ``simJLA'' training sample.  Training on the {\it actual} JLA training sample is also a useful test (Section \ref{subsec:jla_orig_training}), but it does not provide a known truth model to validate the training outputs. For this simJLA test, every simulated SN that goes into the training sample has SALT2.JLA as the ``truth'' model, allowing us to test the consistency of input and output models.

We use the \texttt{SNANA} software \citep{Kessler2009a} to  generate Monte Carlo realizations of SN photometry and spectroscopy mimicking the demographics of the JLA training sample of 420 SNe. \texttt{SNANA} simulations are frequently used to  simulate a random realization for a sample, in order to explore SN distance biases as a function of redshift.Our goal here is different: we aim to accurately simulate {\it every individual} event in the JLA sample, including cadence and signal-to-noise for both photometric and spectroscopic observations. We therefore use the cadence, redshift, and best fit $c$ and $x_1$ from each JLA light curve as input to the simulation. Each set of measured SN properties ($z$, $t_0$, $x_1$,$c$) is used to simulate a rest-frame SED with the SALT2.JLA model. Spectral variations (intrinsic  scatter) are added to the rest-frame SED using the method in K13 and the covariance model from the \citet[hereafter C11]{Chotard2011} scatter model\footnote{We simulate $\alpha=0.14$ and $\beta=3.5$.  Note that the fit value of $\beta$ will be lower than the simulated value by $\sim$0.6 due to the characteristics of the C11 scatter model \citep{Scolnic2016}.}. We do not use the default \citet[hereafter G10]{Guy2010}, as the G10 model has non-physical scatter at extremely red wavelengths that does not match observations. As described in \citet{Kessler2019} the simulation applies cosmological dimming, lensing, peculiar velocity, galactic extinction, and redshifting to produce a redshifted SED at the top of the atmosphere. The filter transmissions and cadence are used to determine measured fluxes and uncertainties. Finally, the noise properties of each measured spectrum are applied to the simulated SED.

\begin{figure}
    \centering
    \includegraphics[width=\columnwidth]{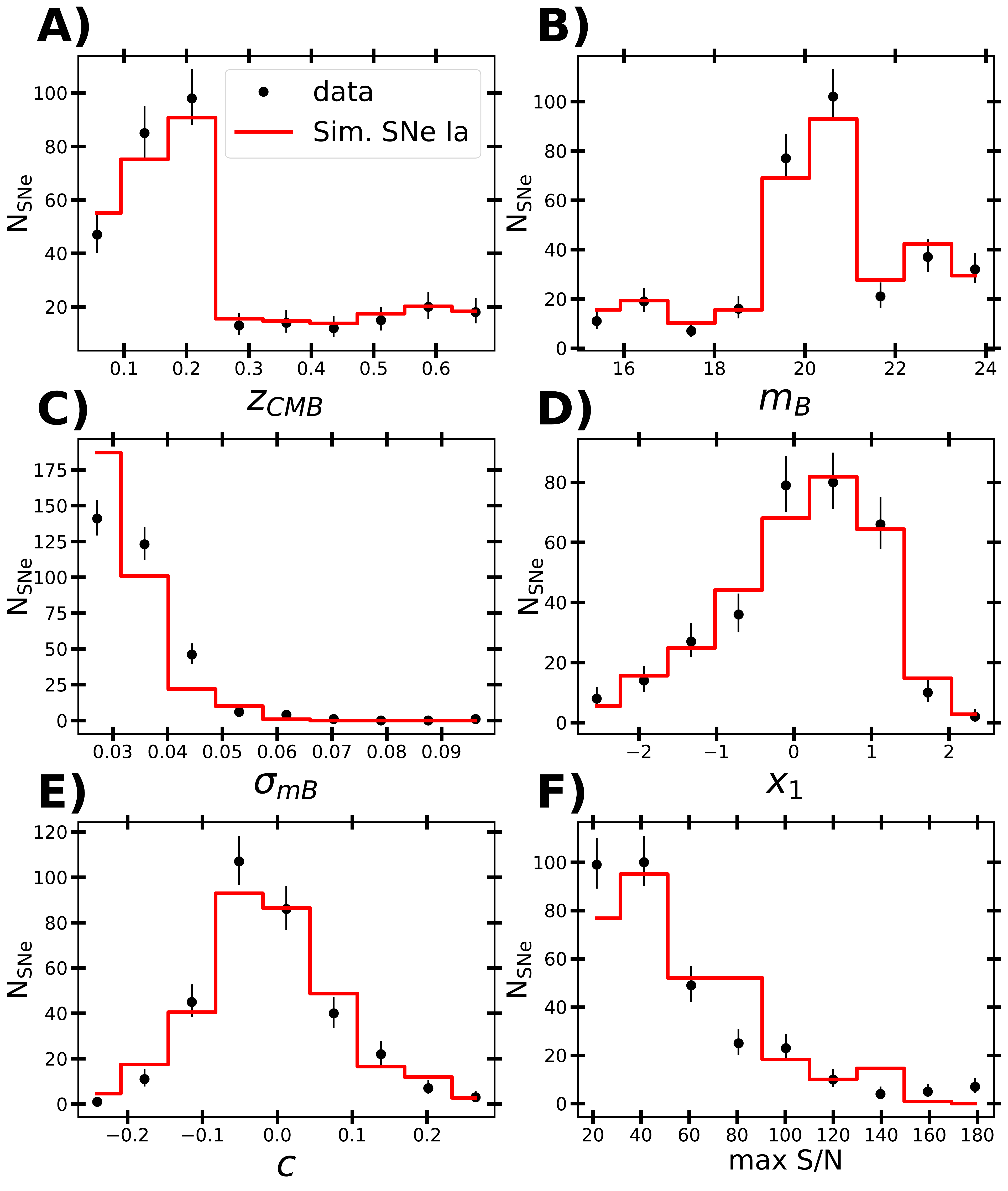}
    \caption{Distributions of parameters for the simulated (red lines) versus real (black dots) SALT2.JLA training samples. Comparisons are shown for redshift (A), $m_B$ (B), uncertainty on $m_B$ (C), $x_1$ (D), $c$ (E), and the maximum S/N in each SN light curve (F).  All distributions are well-matched except for the uncertainty on $m_B$, which is slightly higher in the real data, perhaps due to the C11 model resulting in fainter SNe than the data in some bandpasses.}
    \label{fig:jlatrain_dist}
\end{figure}

This simulation does not exactly reproduce the data set because of random fluctuations in photometric noise and intrinsic scatter, and also because the underlying SALT2 model formalism is an approximation as discussed in \citet{Pierel2020}. Nonetheless the simulation is very similar to the data and is therefore sufficient for testing \texttt{SALTshaker}, as illustrated in Figure \ref{fig:jlasim} for a representative low-$z$ SN\,Ia.  Comparisons between the parameters of simulations and data after fitting with the SALT2.JLA model are shown in Figure \ref{fig:jlatrain_dist}, with only slight observed differences in average $m_B$ uncertainty.

\subsection{Training on a Simulated SALT2.JLA Training Set}
\label{subsec:simjlatraining}

\begin{figure}
    \centering
    \includegraphics[width=3.4in]{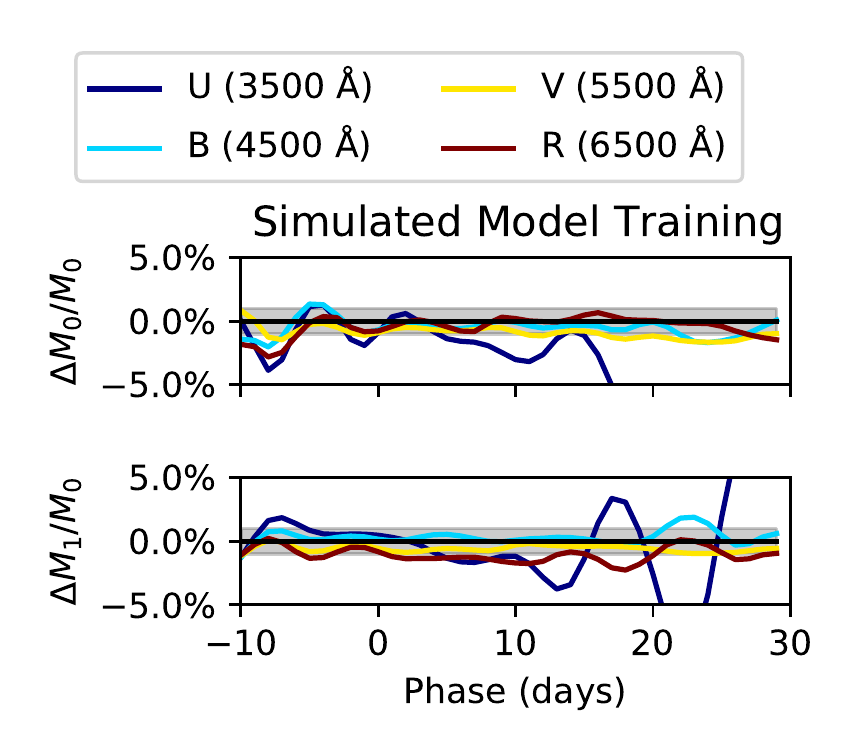}
   \caption{Relative difference between integrated model surfaces of the input SALT2.JLA truth model and \texttt{SALTshaker} model trained on simulated data. Grey shading shows $\pm 1\%$. Lightcurves are integrated over square $UBVR$-like bandpasses $1000$~\AA\ wide, centered at 1000 \AA\ intervals from 3500~\AA\ to 6500~\AA. Significant discrepancies are seen in the ultraviolet where there is sparse  data and regularization schemes have largest impact. }
    \label{fig:lcsimdiffcomp}
\end{figure}

\begin{figure}
    \centering
    \includegraphics[width=3.4in]{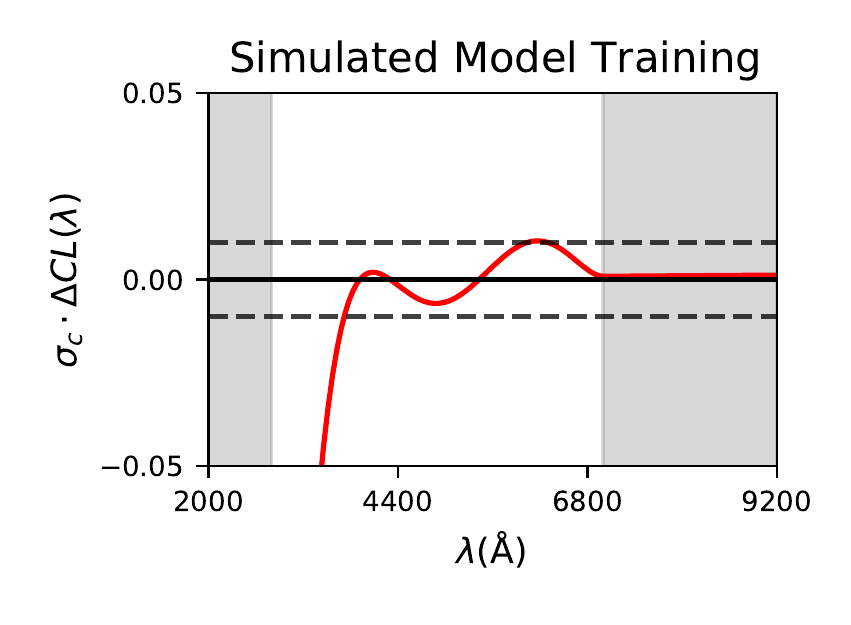}
   \caption{Change between color law of input truth model and model trained on simulated data, multiplied by $\sigma_c=0.1$ to show the impact on a typical SN Ia light curve in magnitudes. Dashed lines show $\pm \SI{.01}{\mag}$. Grey shaded regions indicate the wavelengths for which the color law is linear. The slopes of the color law match closely at $> 7000$ \AA, leading to very small differences across the wavelength range where the color law is linear.}
    \label{fig:colorlawsimdiffcomp}
\end{figure}

We use \texttt{SALTshaker} to train a model using our simulated JLA sample, producing a model we call SALT3.simJLA.  In Figure \ref{fig:lcsimdiffcomp} we show the relative difference between the input SALT2.JLA synthetic photometry and the corresponding  synthetic photometry recovered from the SALT3.simJLA model. 
There is significant discrepancy in the ultraviolet where regularization strongly impacts the recovered model spectra, but at central passband wavelengths 4000~\AA~$<\lambda <7000$~\AA\ and $\SI{-10} {\days} < p < \SI{30} {\days} $ the other three light curves ($B$, $V$, and $R$) are consistent with the truth model at a level better than 1\%. To illustrate the impact of changes in the color law on a ``typical'' light curve in units of magnitudes, we use the quantity $\sigma_c \cdot \Delta CL(\lambda)$, where the standard deviation of the distribution of the SALT color parameter $c$ is $\sim 0.1$. 
As can be seen from Figure \ref{fig:colorlawsimdiffcomp}, the color law difference is $> \SI{0.05}{\mag}$ when $\lambda < 3000 $~\AA, the regime where the color law is least constrained. 

We find that the RMS of $\sigma_c \Delta CL(\lambda)$ is  $\SI{0.01}{\mag}$ between wavelengths $3500\AA < \lambda <7000 $~\AA, central filter wavelengths for which the SALT2.JLA model is considered reliable. We conclude that because few SNe constrain the SED at $\lambda < 3000 $~\AA, the limited set of lightcurve parameters $c$,$x_1$ in this wavelength region poorly constrain the color law and the spectral components. In Sections  \ref{sec:data} and \ref{sec:extended_salt} we substantially expand the training data to address this issue.

\begin{figure*}
    \centering
    \includegraphics[width=7in]{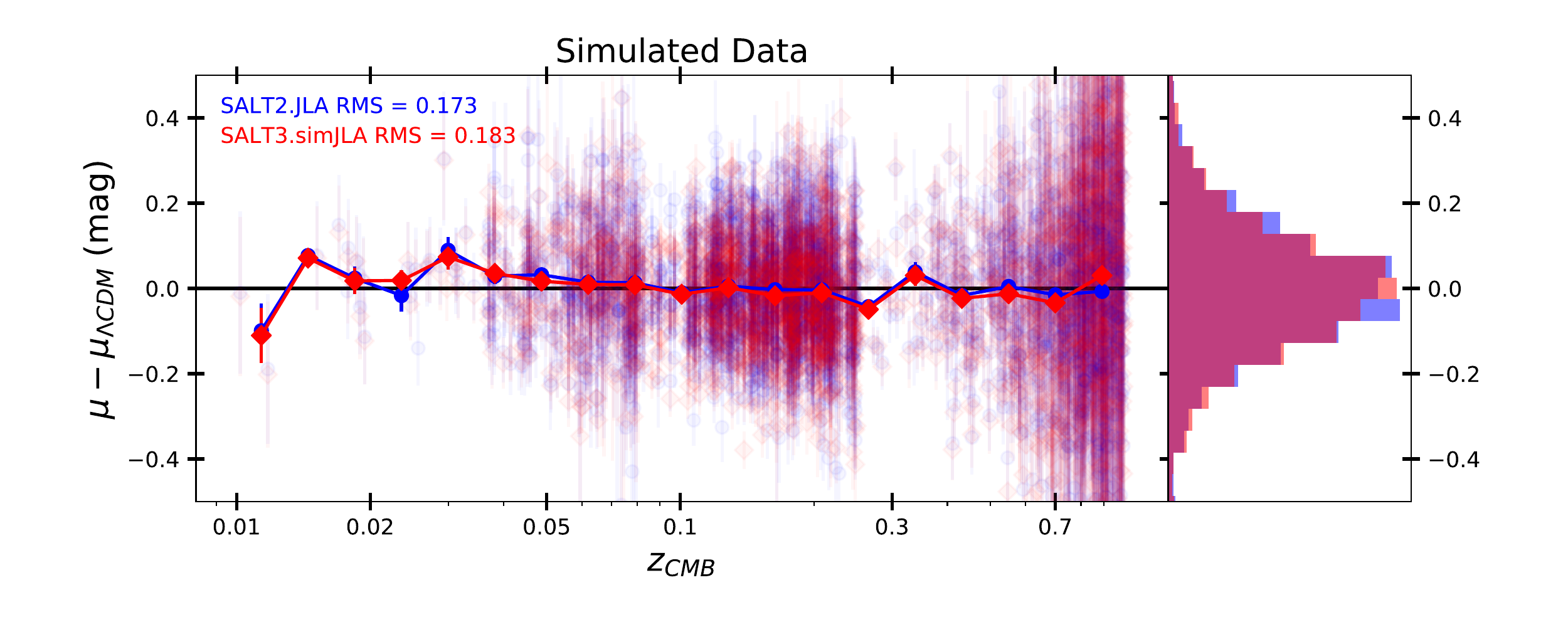}
    \caption{Hubble residual comparison with SALT2.JLA and SALT3.simJLA fits to simulated data.  Each point represents a simulated SN light curve from the simJLA sample, which has been fit with either the SALT2.JLA model (blue) or the SALT3.simJLA model (red) with larger points showing the binned distances from each model.  We do not expect SALT3 to outperform SALT2 here because SALT2 is the ``truth'' model, but we find consistent results.}
    \label{fig:salt3simhubble}
\end{figure*}

Next we compare distances from the two models. To avoid the statistics of the training sample limiting the precision of our validation, we simulate a ``large'' JLA simulation for validation, with summary statistics for the Hubble diagram fits shown in Table \ref{table:saltstats}, row 1.  Instead of simulating exact $x_1$ and $c$ values for an ``apples-to-apples'' comparison with the SALT2.JLA training sample, we  generate $\sim$3000~SNe mimicking a combination of CfA3 \citep{Hicken2009}, SDSS, and SNLS by using $x_1$ and $c$ distributions from \citet{Scolnic2016}. We show an example of Hubble residuals in Figure \ref{fig:salt3simhubble}.  We observe consistent measurements of $\beta$ and RMS$(\Delta \mu)$, with $\sigma_\textit{int}$ higher by \SI{0.005}{\mag} due to our treatment of uncertainties. We also observe a slightly higher value of $\alpha$, which is attributed to the $x_1$/$M_1$ degeneracy. The distance difference between the models (Eq.~\ref{eq:distancediff}) is consistent 
($\textrm{Diff}(\Delta_z \mu) =   \SI{7\pm 11}{\milli\mag}$) despite the U-band discrepancies in the light curves seen in Figures \ref{fig:lcsimdiffcomp} and \ref{fig:colorlawsimdiffcomp}. As the validation sample here has similar demographics to the training sample, the same sparsity of data that allows the observed ultraviolet divergences causes those same divergences to have little effect on the inferred distances. We conclude that for a SALT model to be reliable at these wavelengths, the density of data in this region must be increased to match the density of data where we recover the input model to $\sim 1\%$, at least a factor of two increase (see Section \ref{sec:data}, where we discuss the density of photometric data across the JLA wavelength range).


\subsection{Training on the JLA Training Set}
\label{subsec:jla_orig_training}

\begin{figure*}
    \centering
    \includegraphics[width=7in]{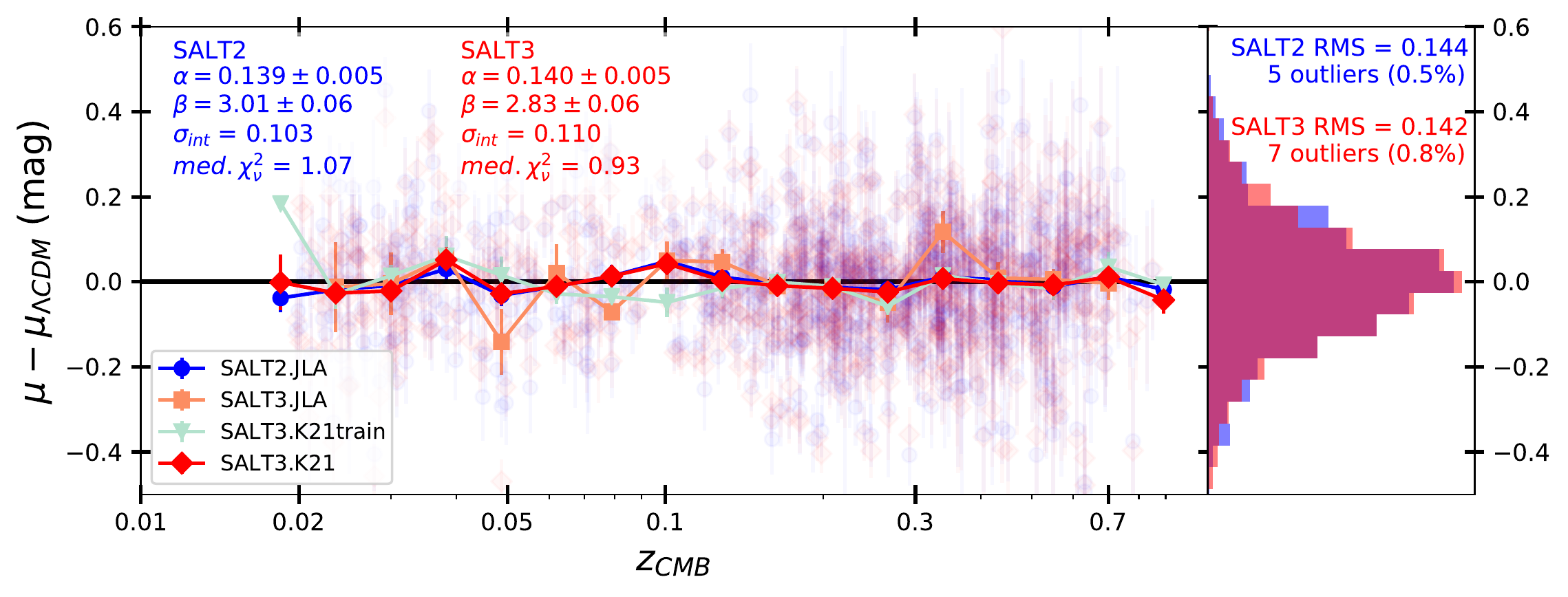}
    \caption{Comparison of the SALT2.JLA (blue) versus SALT3.K21 (red) Hubble residuals for the full K21 compilation, with results from SALT3.JLA and SALT3.K21train shown in orange and teal, respectively. The same number of $3\sigma$ outliers are seen when using either SALT2.JLA and SALT3.K21\textsuperscript{a}. A small change in beta can be attributed to a modified separation of color and stretch, along with improved parameter constraints on color.  Additional details regarding the nuisance parameters in different training sets are shown in Table \ref{table:saltstats}.} 
    \small\textsuperscript{a} SALT3 $3\sigma$ outliers are the SNe 1998ab, 05D2ci, 1995ac, 5635, 40166, 160214, and 370369. SALT2 outliers are the SNe 05D2ci, 40166, 90037, 160214, and 2002hu.
    \label{fig:salt3hubble}
\end{figure*}

Next, we run \texttt{SALTshaker} on the real JLA training sample; we refer to this trained model as SALT3.JLA.  We follow the validation procedure of Section \ref{subsec:validationprocedure} using the JLA training set for our validation sample. As shown in row 2 of Table \ref{table:saltstats}, we find a similar $\alpha$, a slightly lower $\beta$, and consistent RMS$(\Delta \mu)$. We find that $\textrm{Diff} (\Delta_z \mu) = \SI{32 \pm 28}{\milli\mag}$ is consistent with zero. $\sigma_\textrm{int}$  is slightly higher for SALT3.JLA, attributable to greatly decreased model uncertainties. Given the equivalent RMS scatter, this $\sigma_\textrm{int}$ difference does not result in reduced distance precision. The distance moduli are consistent at the 1-$\sigma$ level between SALT2.JLA and the SALT3.JLA, and the model surfaces are consistent within the uncertainties.  We show the Hubble residuals of this model in Figure \ref{fig:salt3hubble} (along with our extended SALT3 models discussed in Section \ref{sec:extended_salt}). 

\section{Expanded Training Data: The K21 Compilation}
\label{sec:data}

\begin{figure}
    \centering
    \includegraphics[width=3.4in]{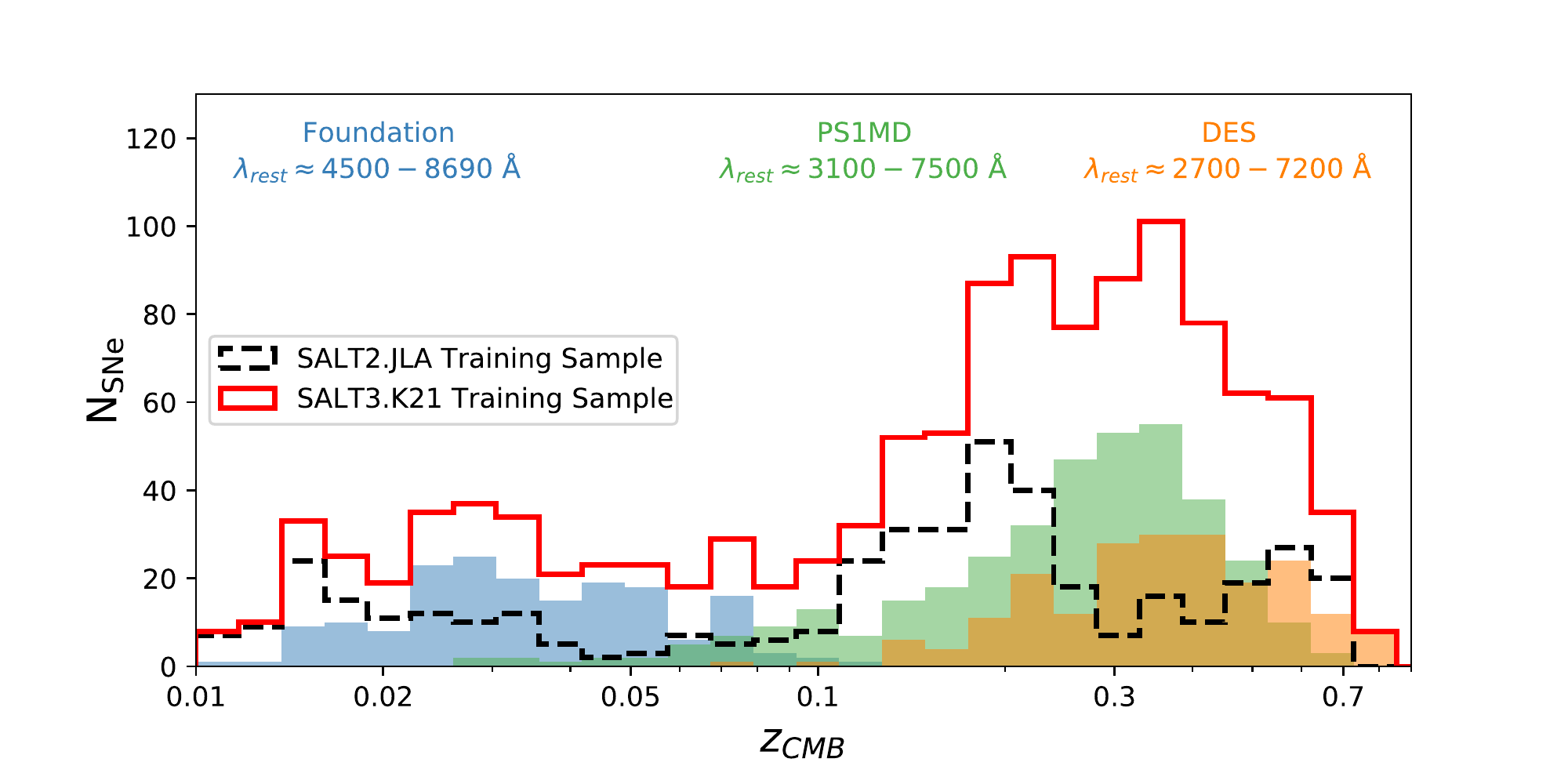}
    \caption{Redshift distribution of the JLA training data for SALT2.JLA (black dashed) and our training data for SALT3.K21 (red solid), which adds Foundation (blue), PS1 (green) and DES (orange) to the JLA training sample as well as small samples from CSP and CfA4. At the top, we label the rest-frame central filter wavelengths of each added dataset, with the full filter widths in conjunction with spectroscopy constraining the full SALT3.K21 wavelength range.  The combined dataset contains approximately three times as many spectra as the previous SALT2 training sample and contains approximately 2.5 times as many SNe.}
    \label{fig:redshifts}
\end{figure}

\begin{figure*}
    
    \includegraphics[width=7in]{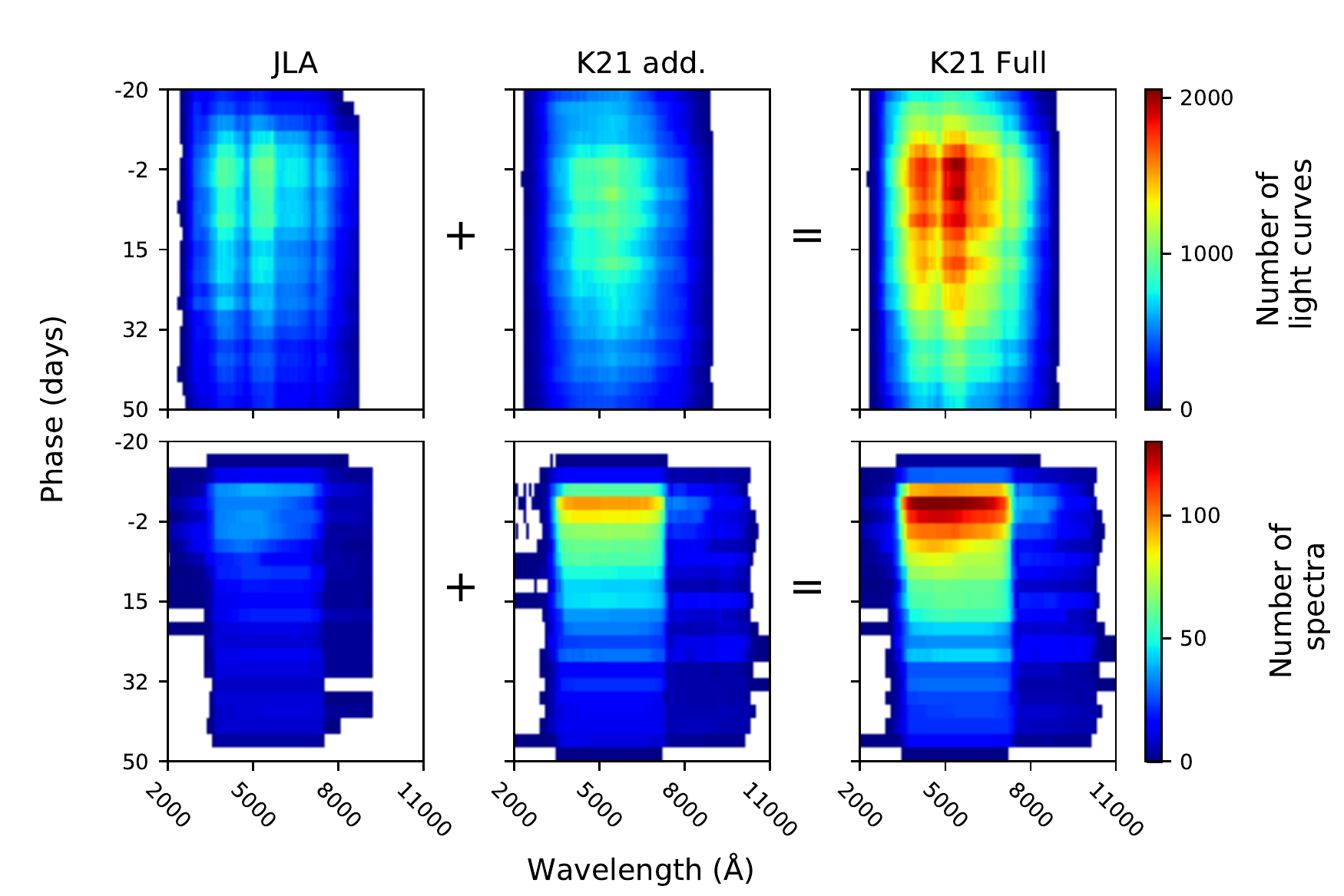}
    \caption{Number of light curves or spectra constraining each bin in our phase/wavelength space. Panels show coverage of photometric data (upper panels) and spectral data (lower panels) in the JLA training sample (left panels), the PS1, DES, and Foundation data added in the K21 compilation (middle panels), and the complete K21 compilation (right panels). A bin is considered ``covered'' by a given light curve if it is within the FWHM of the rest-frame bandpass. Foundation photometry provides much more extensive photometric coverage of red wavelengths, while PS1 and DES photometry provides additional rest-frame blue photometry, and the new photometric data covers the phase space more uniformly than JLA alone. Similarly, spectra from \texttt{kaepora} are immensely impactful at wavelengths $\lambda>7500$ \AA, greatly assisting the deconvolution of $z$ band data.}
     \label{fig:coveragecomparison}
\end{figure*}

To create a next-generation SALT model with extended wavelength range and reduced uncertainties, we add high-quality data from the Dark Energy Survey \citep{Brout2019b}, the Foundation Supernova Survey \citep{Foley2018}, and the Pan-STARRS Medium Deep Survey \citep{Scolnic2018}. For SNe with photometric data in the training sample, we add 693 low-$z$ spectra from the Kaepora database \citep{Siebert19}, the majority of which originate from the Berkeley SN\,Ia Program (BSNIP; \citealp{Silverman12,Stahl20}). We show the density of photometric and spectral data in phase and wavelength space with the original JLA training sample and the additional data included in the K21 compilation in Figure \ref{fig:coveragecomparison}. Our final training sample adds data across the phase space, but is most impactful in the red and blue regions of our wavelength range, where the JLA training data were limited. Wavelengths $<3500$ \AA\ are on average covered by  $1.8 \times$ more light curves, while wavelengths $>7500$ \AA\ are covered by an average of $2.1 \times$ as many light curves and $5.7 \times$ as many spectra.  Additional photometric data makes the distribution across phase space more uniform, where the JLA data has comparatively little data in the gap between low-redshift $B$ and $V$ bandpasses. 

We initially characterize our performance with the extended set of data using separate samples for training and validation. We define the ``K21valid'' and ``K21train'' compilations by randomly assigning half of the supernovae to each. Finally, to produce our best model,  we combine all available data to create the ``K21'' compilation, and summarize the data included below and in Table \ref{table:trainingsamp}.  The redshift distributions of these data are shown in Figure \ref{fig:redshifts}. Trained models using these compilations as training samples are discussed in Section \ref{sec:extended_salt}.


\subsection{The Pan-STARRS Medium Deep Survey}

The Pan-STARRS medium deep survey covered 70 square degrees of sky over four years, discovering approximately 5200 SNe \citep{Jones2017,Villar2020} and spectroscopically classifying $\sim$10\% of these at a median redshift of $\sim$0.35 \citep{Rest2014}.  The Pantheon analysis, which combined these data with JLA, includes 279 PS1-observed SNe\,Ia with an average of approximately 6 observations per 10 days in $griz$ \citep{Scolnic2018}. We use these \sne\ in our training data.

\subsection{The Foundation Supernova Survey}

The Foundation Supernova Survey \citep{Foley2018} followed SNe using the Pan-STARRS1 telescope, and measured well-calibrated SN light curves in $griz$ filters with a five-day cadence near maximum light and an approximately 8-day cadence beginning at $+$10 days after maximum light.  To achieve reduced selection effects compared to previous surveys that targeted bright, pre-selected low-$z$ galaxies, Foundation primarily followed SNe discovered by untargeted surveys such as the All-Sky Automated Survey for Supernovae \citep{Shappee2014ASASSN}, the Asteroid Terrestrial-impact Last Alert System \citep{Tonry2018ATLAS}, Gaia \citep{Gaia2016Reference}, and the Pan-STARRS Survey for Transients \citep{Huber2015PSST} .

The Foundation first data release in \citet{Foley2018} contains 225 SNe\,Ia, 180 of which are cosmologically useful.  These data have been used to measure cosmological parameters in \citet{Jones2019} and the correlation of host galaxy properties with SN distances in \citet{Jones2018}.  The $iz$ band coverage of Foundation is particularly critical to creating a SALT3 model that is trained to redder wavelengths than enabled by the JLA data alone.  We include spectra for 114 Foundation SNe from the \citet{Dettman2021PhotosphericVelocity} data release.

\subsection{The Dark Energy Survey}

The Dark Energy Survey (DES) three-year spectroscopically classified SN sample contains 207 SNe\,Ia at a median redshift of 0.36 \citep{Abbott2019}.  These SNe were discovered by imaging eight 2.7 deg$^2$ ``shallow'' fields (depth $\approx 23.5$~mag) and two 2.7 deg$^2$ ``deep'' fields (depth $\approx 24.5$~mag) approximately once per week \citep{Smith2020}.  Transients were discovered using a difference-imaging pipeline \citep{Kessler2015} and final photometry was performed with a ``scene modeling'' pipeline described in \citet{Brout2019b}.  See \citet{Abbott2019} and \citet{Brout2019} for additional details regarding the DES SN\,Ia data and analyses.  These data have a maximum redshift of $\sim$0.85 and complement SNLS in probing rest-frame near-UV wavelengths with well-calibrated (sub-percent) photometric data \citep{Burke2018DESCalibration}.

\begin{table*}
    \centering
    \caption{The K21 Compilation}
    \begin{tabular}{lrrrrrrrr}
        \hline \hline\\[-1.5ex]
        Survey & N$_{\mathrm{SN}}$ &N$_{\mathrm{spectra}}^{\rm a}$&N$_{\mathrm{spectra}}^{\rm a}$ & $z_{min}$&$z_{med}$&$z_{max}$&Filters&Ref.\\
        &&(JLA)&(Total)&&&&&\\
        \hline\\[-1.5ex]
Calan-Tololo&5&0&0&0.015&0.020&0.051&$BVRI^{\rm b}$&\citet{Hamuy1996}\\
CfA1&8&46&66&0.004&0.011&0.050&$UBVRI^{\rm b}$&\citet{Riess1999}\\
CfA2&13&108&166&0.008&0.014&0.031&$UBVRI^{\rm b}$&\citet{Jha2006}\\
CfA3$^{\rm c}$&51&31&534&0.004&0.023&0.041&$UBVRIri$&\citet{Hicken2009a}\\
SDSS$^{\rm c}$&202&0&12&0.037&0.166&0.250&$ugriz$&\citet{Holtzman08}\\
SNLS&111&63&63&0.149&0.499&0.700&$griz$&\citet{Astier2006}\\
Misc. low-$z^{\rm c}$&25&152&216&0.001&0.009&0.077&$UBVRI^{\rm b}$&\citet{Jha2007}\\
{\bf SALT2.JLA Total}&415&400&1057&0.001&0.172&0.700&\nodata&\nodata\\
\hline
CfA4$^{\rm c}$&30&0&0&0.009&0.029&0.070&$BVri$&\citet{Hicken2012}\\
CSP&13&0&36&0.011&0.029&0.058&$uBVgri$&\citet{Krisciunas2017}\\
Foundation&153&0&114&0.005&0.034&0.111&$griz$&\citet{Foley2018}\\
PS1 MDS$^{\rm c}$&266&0&0&0.026&0.297&0.630&$griz$&\citet{Scolnic2018}\\
DES-spec&206&0&0&0.078&0.362&0.850&$griz$&\citet{Abbott2019}\\
{\bf New Data Total}&668&0&150&0.005&0.244&0.850&\nodata&\nodata\\
\hline\\[-1.5ex]
{\bf SALT3 Total}&1083&400&1207&0.001&0.201&0.850&\nodata&\nodata\\
        \hline\\[-1.5ex]
        \multicolumn{9}{l}{
        \begin{minipage}{6in}
        {\bf Note.} SNe\,Ia in the K21 compilation, after cuts: the original JLA training sample is reduced to 415 SNe (originally 420) and the new data adds 668 SNe, for a total of 1083 SNe.
        
        $^{\rm a}$ Total number of spectra, rather than number of SNe with spectra, which is 83 for the JLA sample and an additional 297 from the new data included here.  Spectra are from \citet{Filippenko1992,Wells1994,Patat1996,NATO1997,Li2001,Salvo2001,Valentini2003,Anupama2005,Benetti2004,Kotak2005,Leonard2005,Garavini2007,Stanishev2007,Thomas2007,Foley2008,Pignata2008,Wang2009,Foley2010,Ostman2011,Walker2011,Blondin2012,Silverman2012,Folatelli2013,Balland2018} and from private communication with M.\ Betoule and C.\ Balland.
        
        $^{\mathrm{b}}$ We note that these filter transmission curves were not provided for these samples.
        
        $^{\mathrm{c}}$ 11 SNe in these samples have additional data from the CSP \citep{Contreras2010,Krisciunas2017}, while SDSS SNe 2006oa, 2006ob, 2006on, and 2006nz also have data from CfA3 \citep{Hicken2009a}.
        
    $^{\mathrm{c}}$ See also \citet{Rest2014} and \citet{Jones2018}.
    
    \end{minipage}}
    \end{tabular}
    \label{table:trainingsamp}
\end{table*}

\subsection{Sample Selection Cuts}
\label{sec:cuts}

To ensure that SNe are suitable for inclusion in a training sample, we first require that every SN in the compilation is a spectroscopically classified, Branch-normal SN\,Ia  (\citealp{Branch1993Normal}; we include 1991T-like SNe\,Ia following the original SALT2 training procedure; \citealp{Guy2007}).  For the SN light curves, we make the following selection requirements (cuts):

\begin{itemize}
    \item At least four epochs at phases between $-10 <p<35$ days, where $p$ is the rest-frame phase relative to time of B-band peak. 
    \item At least one measurement after peak brightness ($5 < p < 20$), to constrain the shape.
    \item At least one measurement in each of at least two filters at $-8 < p < 20$ to constrain the color of the SN.
    \item At least one measurement prior to peak brightness ($-10 < p < -1$) to ensure a well-measured time of maximum light.  This is the only cut that was not included in the original SALT2 training\footnote{5 of 420 SNe from the JLA sample are removed with this cut.}.
\end{itemize}

For the spectra, we include the original SALT2.JLA training spectra in addition to spectra taken from the \texttt{kaepora} database and spectra taken as part of the Foundation Supernova Survey.  A number of these spectra were taken by the Foundation team, but most have been published on the Transient Name Server\footnote{\url{https://wis-tns.org/}} as classification spectra.  

To ensure minimal host galaxy contamination in the high-$z$ SN spectra used for model training, \citet{Guy2007} fit the spectra with a combined model including the predicted SN spectrum, the spectral recalibration parameters, and a galaxy model (elliptical, S0, Sa, Sb, and Sc templates).  They removed spectra for which there was evidence for host galaxy contamination at the 68\% confidence level.

For Foundation and the additional low-$z$ SN spectra included here, low-$z$ SNe are much brighter relative to their host galaxies than at the redshifts probed by SNLS and SDSS.  We clip host galaxy lines, mask regions with uncorrected telluric features, and remove excessively noisy or poorly calibrated regions of each spectrum, but do not attempt to subtract a host galaxy continuum.  We remove a handful of spectra with poor quality from visual inspection.  After cuts there are 114 spectra from Foundation and 693 from Kaepora, a subset of which are shown in Section \ref{sec:extended_salt}. All but five Foundation SNe have redshifts measured from host galaxy features. 
The complete training data are available at \url{https://saltshaker.readthedocs.io}.

\section{The SALT3 Model: Extending the Wavelength Range and Training on Pantheon, Foundation, and DES Data}
\label{sec:extended_salt}

Having demonstrated the effectiveness of \texttt{SALTshaker} on the original JLA training data, we now train a SALT3 model using additional data, extending the free model parameters in \texttt{SALTshaker} in three ways.  First, we extend the SED wavelength range to 11000~\AA\ so that rest frame filters centered at wavelengths up to $\sim 8500$ angstroms, such as the Foundation $z$-band, can be fit with the model. 
Secondly, we extend the SED wavelength range over which the color law is fit with a polynomial (Eq. \ref{eq:colorlaw}) to 8000~\AA\ , just below the central wavelength of the PS1 $z$ band.  We use a fourth-order polynomial, rather than the third-order polynomial used for SALT2.JLA, to model the color law over the increased wavelength range with $\times 2.5$ more training data. Third, the color scatter model is changed from a third to a fourth-order polynomial to allow additional flexibility over the increased wavelength range.

\subsection{Validation on Extended Wavelength Range}
\label{subsec:separatetrainvalid}
Before presenting our model trained on all data in Section \ref{subsec:completesampletraining}, we check the performance on separate training and validation samples. We take the ``K21valid'' compilation as our cosmology sample using our ``SALT3.K21train'' model as described in Section \ref{subsec:validationprocedure}, and show the resulting Hubble residuals in Figure \ref{fig:salt3hubble} and in row 3 of Table \ref{table:saltstats}.  We find that nuisance parameters are similar between SALT2.JLA and the SALT3.K21train model, with $\sigma_\textrm{int}$ slightly higher by \SI{0.01}{mag} but with a consistent $RMS(\Delta \mu)$ (the SALT3 RMS is negligibly smaller).  $\textrm{Diff} (\Delta_z \mu)$ is consistent with zero at the mmag level.

\begin{figure*}
    \centering
    \includegraphics[width=7in]{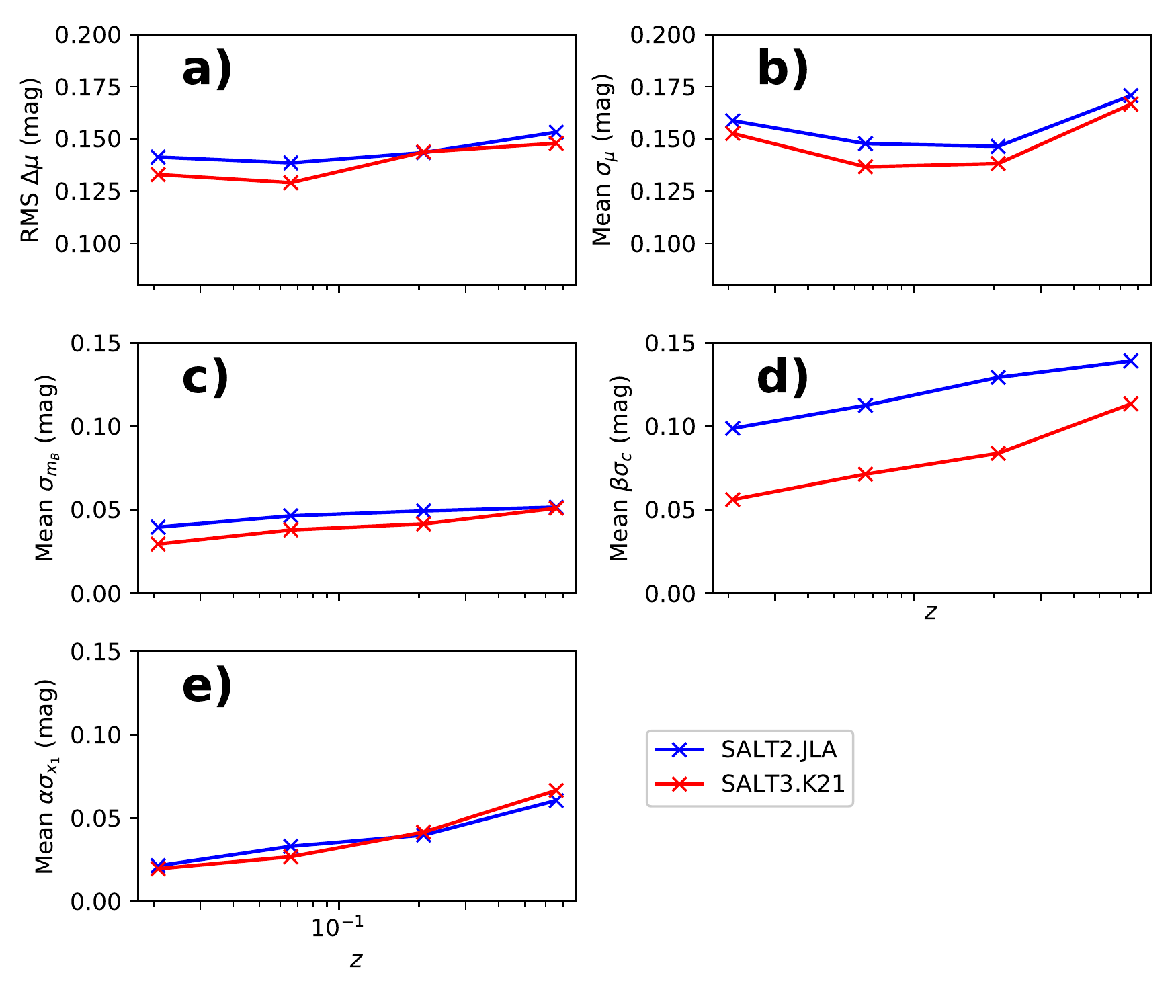}
    \caption{RMS of Hubble residuals (panel a) and quadratic means of uncertainties in distances and light-curve parameters (panels b-e) of our compilation using two SALT models in four logarithmically spaced redshift bins from $0.015<z<1$. SALT3.K21 improves measured RMS Hubble scatter across our entire redshift range. The smallest improvements appear to be in the intermediate redshift range $z\sim 0.2$, with larger improvements at both lower and higher redshifts where improvements to the red and blue regions of the rest-frame model are most important.
    }
    \label{fig:errorstatistics}
\end{figure*}

\subsection{Training on Complete K21 Compilation}
\label{subsec:completesampletraining}
\begin{figure*}
    \centering
    \includegraphics[width=7in]{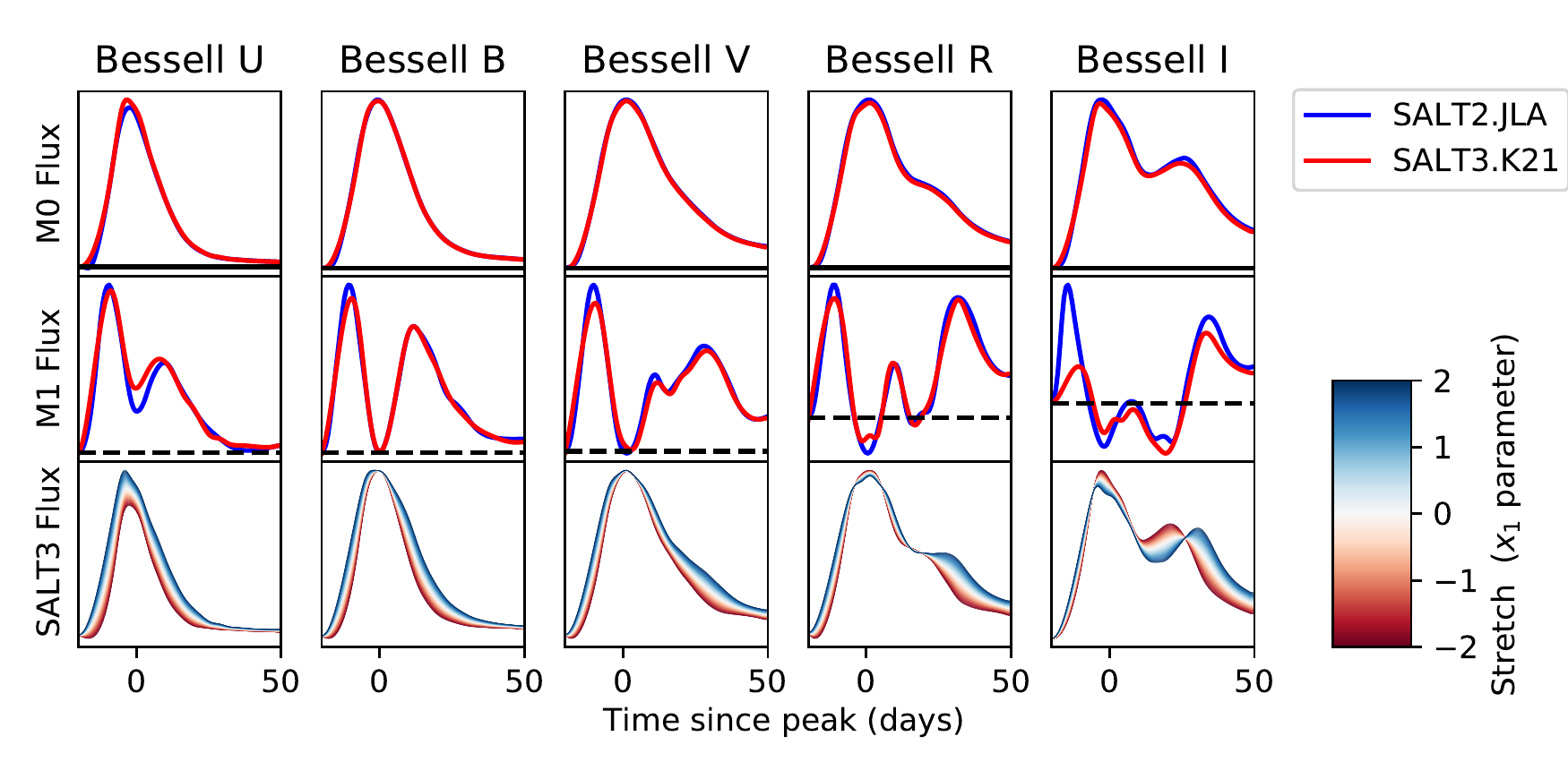}
    \caption{Comparisons of synthetic light curves. Upper two panels show light curves from both SALT2.JLA and SALT3.K21 in arbitrary units of flux. Differences are most prominent in Bessell I band. Lower panels show a family of SALT3 lightcurves created by varying the $x_1$ parameter.} 
    \label{fig:lightcurvesx1}
\end{figure*}

\begin{figure*}
    \centering
    \includegraphics[width=7in]{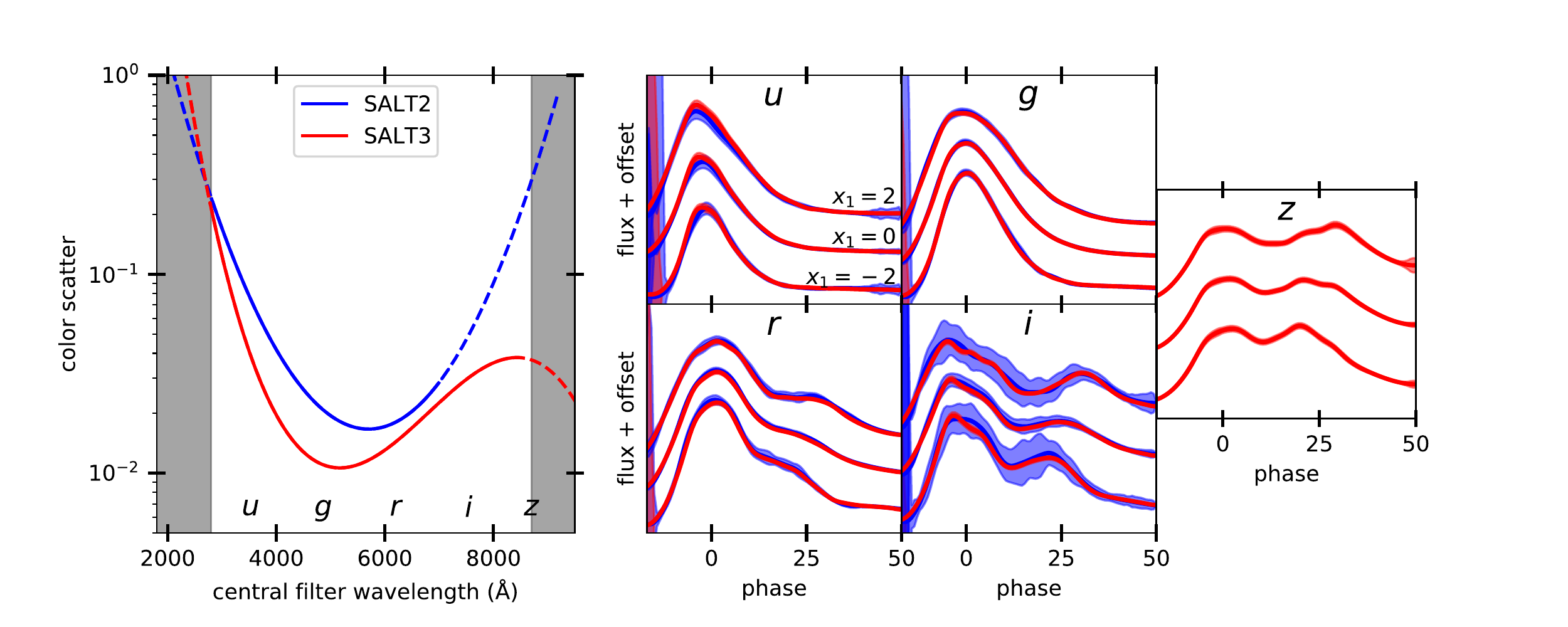}
    \caption{Color scatter as a function of central filter wavelength (left) and example light curves (right) with model uncertainties from SALT2.JLA (blue) and SALT3.K21 (red).  SALT3 has comparable errors in the $u$ and $g$ bands but much smaller uncertainties in the $r$ and $i$ bands due to training data with much better coverage at those wavelengths, especially at large or small $x_1$ (with errors in both models blowing up at very early phases).  Only the SALT3 model covers the $z$ band (right-most panel).  On the left, dashed lines illustrate the wavelengths where the color scatter is unconstrained by data from the JLA sample (blue) or the K21 compilation (red; grey shading).}
    \label{fig:errors}
\end{figure*}

\begin{figure}
     \centering
     \includegraphics[width=3.5in]{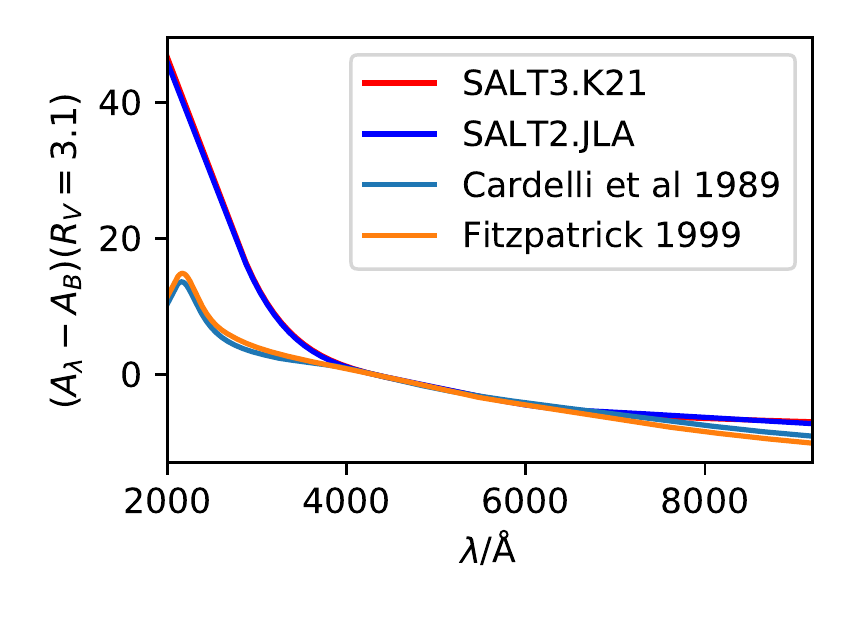}
     \caption{Comparison of SALT2.JLA versus SALT3.K21 color law, along with the extinction curves of \citet{Cardelli99} and \citet{Fitzpatrick99} for comparison. For all four curves $R_V$ is fixed to 3.1. }
     \label{fig:extinctionlawcomparison}
 \end{figure}

Finally, we train our best SALT3 model, which we call SALT3.K21, using all of the data described in previous sections as a training sample. The sample includes 1083 SNe, a factor of 2.5 more SNe than the JLA training sample, and 1207 spectra, a factor of three increase in the number of spectra.  
Synthetic light curves from SALT3.K21 and SALT2.JLA are compared in Figure \ref{fig:lightcurvesx1} and the model uncertainties are compared in Figure \ref{fig:errors}.  
We see good consistency between SALT2.JLA and SALT3.K21, with modest differences in the $u$-band and some additional differences in redder bands; at both wavelength ranges we have substantially increased the training sample (see Figure \ref{fig:coveragecomparison}). 
Similarly, as shown in Figure \ref{fig:extinctionlawcomparison}, the color law is consistent with that of SALT2.JLA to within 1\% across the entire wavelength range.

We compare Hubble residuals of the SALT3.K21 and SALT2.JLA models in Figure \ref{fig:salt3hubble}. Individual standardized distances are consistent to $\SI{.05}{\mag}$ between the two models, and the effects on the Hubble diagram are found in row 4 of Table \ref{table:saltstats}. $\textrm{Diff} (\Delta_z \mu)$ is consistent with zero at  \SI{2\pm 14}{\milli\mag}. Finally, row 4 of Table \ref{table:saltstats} shows nuisance parameters and Hubble diagram metrics, demonstrating that \texttt{SALTshaker} produces a new SALT3 model with slightly lower total dispersion, consistent $\sigma_\textrm{int}$, and consistent distances.  We note that the $\beta$ parameter is lower by 0.24 in the SALT3 model, likely due to a reconsidered separation of color and stretch.

\subsubsection{Model Uncertainties and Hubble Scatter of SALT3.K21}
The uncertainties in Figure \ref{fig:errors} show that both the color scatter and light-curve uncertainties in redder bands are much lower in SALT3.K21 than SALT2.JLA. The SALT2 color scatter was effectively unconstrained by the JLA data past mean passband wavelengths $\sim8000 $~\AA, and we find that while the color scatter is significant at these wavelengths, it is much smaller than implied by the SALT2.JLA model. Our additional data constrains this region up to $\sim 8500 $~\AA, and redder data will be required to see how this effect carries into the NIR. We also note lower color scatter by $\sim 1\%$ in the blue, which we attribute to improved relative calibration of the training sample. 

Fitting the light curves using the SALT3.K21 model, we compare the uncertainties in distances and light-curve parameters to those found using SALT2.JLA. In Figure \ref{fig:errorstatistics} we compare performance across the redshift range. Our model shows reduced Hubble scatter and distance uncertainties over the SALT2.JLA model at nearly every redshift, with the least improvement at moderate redshifts $z \sim 0.2$, where signal to noise is high and the SALT2.JLA model is already performing well. Light-curve parameters from SALT3.K21train have smaller uncertainties across the redshift range, with the exception of $x_1$ uncertainties. Our largest improvements, particularly in color uncertainty, are found at low redshift, where the improved red wavelength coverage of our model allows the use of additional light curve bands in fitting these SNe, providing stronger constraints.

\begin{figure}
    \centering
    \includegraphics[width=3.4in]{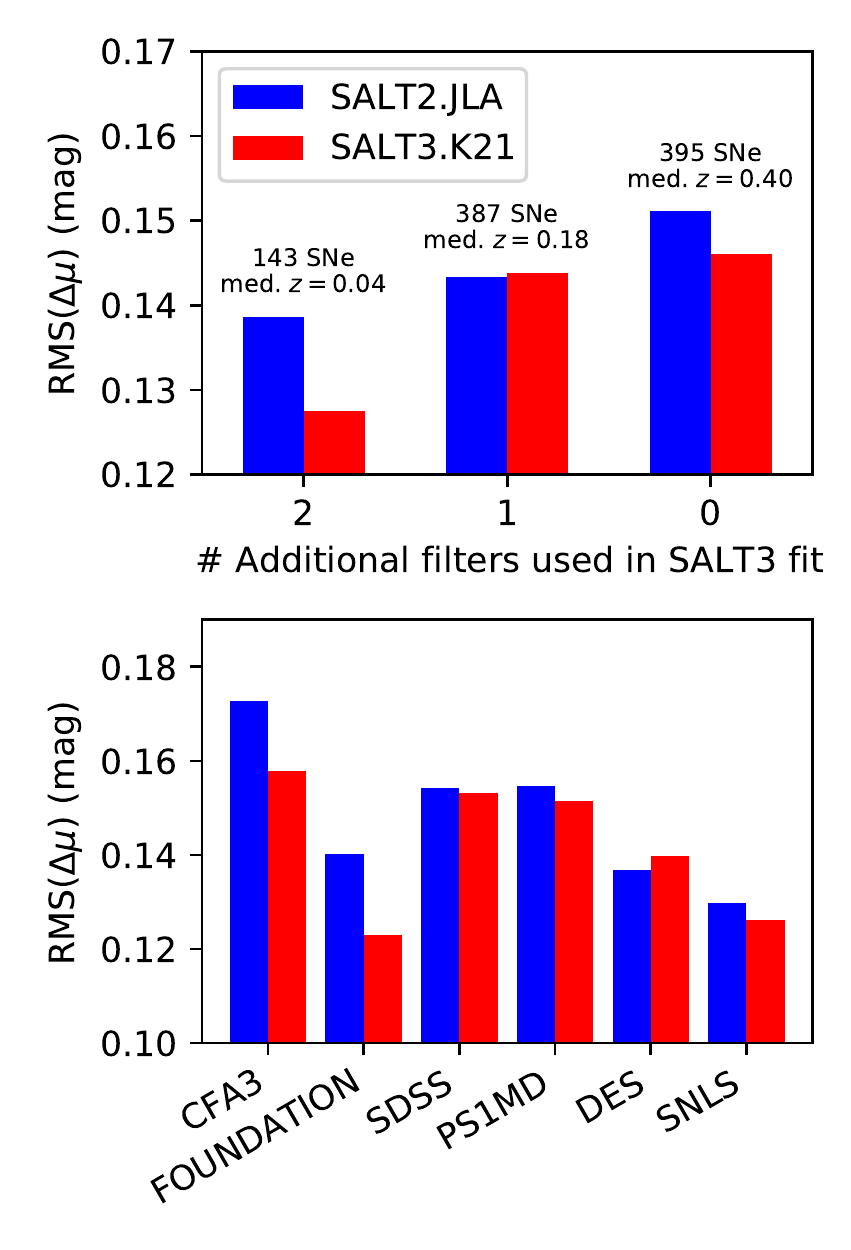}
    \caption{Upper panel: RMS$(\Delta \mu)$ versus the number of additional filters used in SALT3.K21 (red)  light-curve fits as compared to fits made using SALT2.JLA (blue). The number of SNe and median redshift are printed above each bin. Lower panel: RMS$(\Delta \mu)$ versus sample for SALT3.K21 (red) and SALT2.JLA (blue), for all samples which have more than 30 \sne\ in our compilation. As we move to surveys dedicated to searches at higher redshift, fewer SNe have data that could not be fit with SALT2, and the advantage of SALT3's wavelength range becomes less important.}
    \label{fig:scatterbyaddtlfilters}
\end{figure}  

There is a slight improvement in RMS Hubble residuals, largely attributable to decreased color uncertainties in the low redshift sample. Breaking this down further, Figure \ref{fig:scatterbyaddtlfilters} shows the Hubble scatter binned by the number of additional filters used in the SALT3.K21 light-curve fit as compared to the SALT2.JLA light-curve fit in Figure \ref{fig:scatterbyaddtlfilters}. Where two additional filters are available, RMS$(\Delta \mu)$ improves by $\sim 10\%$. We conclude improvement in the SALT3 model is most noticeable when it allows us to fit SNe with existing light curves in filters out of the SALT2.JLA wavelength range. At high redshift, SALT3's improved constraints on the NUV model reduce Hubble scatter by $\sim 0.01$ mag, although there is typically not sufficiently red data to take full advantage of the extended wavelength range at these redshifts. Similarly, when breaking down results by survey, as we show in Figure \ref{fig:scatterbyaddtlfilters}, the greatest improvement is in the low-$z$ CFA3 and Foundation samples where we can make use of $I$ and $iz$ band observations (respectively) previously unused in SALT2.JLA-based analyses at low redshift. We are able to reduce the RMS$(\Delta \mu)$ of the Foundation sample from $0.144 \pm 0.001$ mag to 0.125 mag, an improvement of 15\%.



\subsubsection{Comparison of SALT3.K21 and SALT2.JLA Light-curve Parameters}
\label{subsec:lcparcomparison}
 The \texttt{SALTshaker} training procedure results in different distributions of light-curve parameters compared to SALT2.JLA. Some differences are due to changes in how the SALT3 model is defined, while others are from the demographics of the training sample. For example, a $x_1$/$M_1$ degeneracy is broken by setting a constraint on the standard deviation $\sigma_{x_1} = 1$, both in our procedure and in the original training procedure; therefore including additional data with higher stretch increases the scale of the $M_1$ component. 

\begin{figure}[b!]
    \centering
    \includegraphics[width=3.4in]{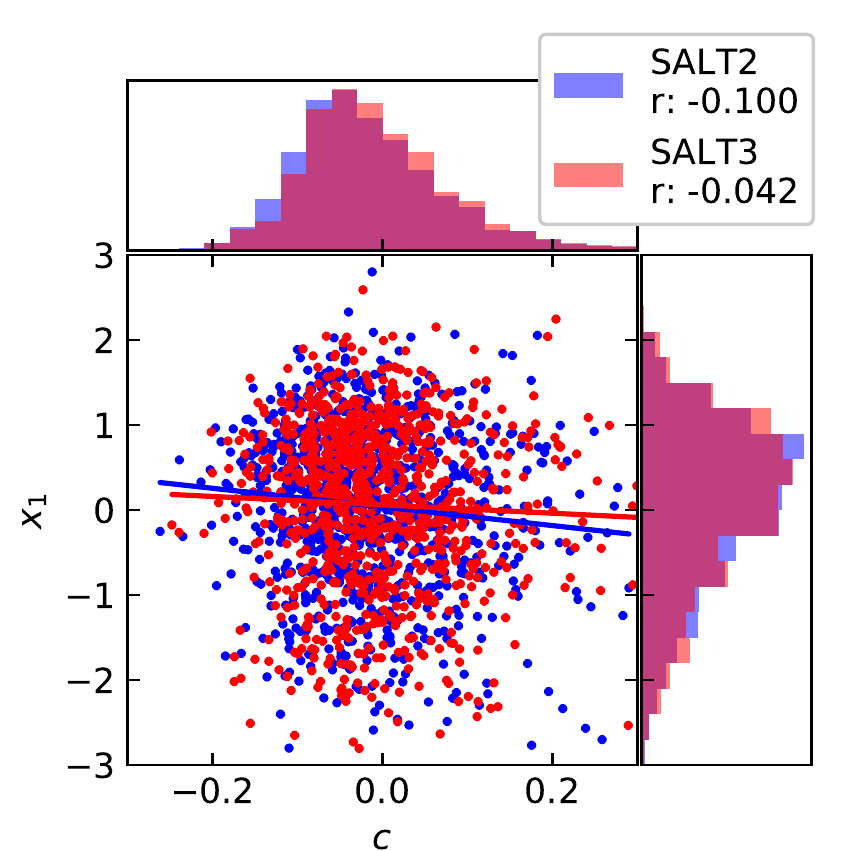}
    \caption{Scatterplot of the $x_1$ and $c$ light-curve parameters measured with SALT2.JLA and SALT3.K21. We observe a linear transformation of the parameters, as expected from changes in the demographics of the training sample and our procedure for separating stretch and color. Best fit $c$-$x_1$ lines and correlation coefficients $r$ are shown to illustrate the rotation of the distribution as SALT3 forces the independence of $x_1$ and $c$.}
    \label{fig:parametercomparison}
\end{figure}

In Figure \ref{fig:parametercomparison} we compare the distributions of the $x_1$ and $c$ parameters from both models. We note a slight rotation, shift, and scale change of the color/stretch distribution relative to SALT2.JLA. These linear transformations cancel in the Tripp standardization of distance, absorbed into the nuisance parameters $\alpha, \beta, \mathcal{M}$. The offset in color in particular is from including bluer high redshift SNe from PS1 and DES in the training sample. The rotation is due to the distinct procedures for separating stretch and color between the SALT2 training code and \texttt{SALTshaker}. SALT3.K21 $x_1$ and $c$ values can be approximated from the SALT2.JLA values by the transformations 

\begin{align}
    x_1^{(\text{SALT3})} \approx & 1.028 x_1^{(\text{SALT2})} + 0.138 c^{(\text{SALT2})}+0.005 \\
    c^{(\text{SALT3})}\approx & 0.002 x_1^{(\text{SALT2})} + 0.985 c^{(\text{SALT2})} +0.013.
\end{align} 

To test the similarity of the shape of the distributions, we use the two dimensional Kolmogorov-Smirnov -like test statistic of \citet{Press19882dkstest}. We modify the procedure by linearly transforming both sets of light-curve parameters to have the same mean, standard deviation, and $x_1$-$c$ correlation before calculating the statistic. We find the test statistic for the SALT2.JLA and SALT3.K21 $x_1$ and $c$ parameters, then bootstrap resample the data and calculate the test statistic with the resampled data. We derive a p-value$=0.895$, and conclude that we cannot distinguish the shape of the underlying distributions.


\setlength{\tabcolsep}{0mm}
\begin{table*}[t]
    \centering
    \caption{Summary of SALT Model Nuisance Parameters and Relative Distance Difference for Different Training Sets}
    \begin{tabular}{l|l r r r r r r r}
    \hline
    \hline
     \multicolumn{4}{c}{Sample}&\multicolumn{3}{c}{Nuisance Parameters}&\multicolumn{2}{c}{Hubble Flow Dist.}\\
     \multicolumn{2}{c}{Model}&$N_\text{Training}$&Validation Sample&$\alpha$&$\beta$&$\sigma_{int}$&RMS$(\Delta \mu)$& $\textrm{Diff} (\Delta_z \mu)$\\
    \hline
&\ SALT2.JLA&420&large sim.\ JLA$^{\rm b}$&$0.151 \pm 0.003$&$3.101 \pm 0.031$&$0.098$\,\,\,&0.173&\nodata\\
\rownumber\  & \ SALT3.simJLA&420&large sim.\ JLA$^{\rm b}$&$0.165 \pm 0.003$&$2.854 \pm 0.028$&$0.103$\,\,\,&0.182&\,\,$-0.007 \pm 0.011$\\
\hline
&\ SALT2.JLA&420&JLA&$0.131 \pm 0.009$&$3.230 \pm 0.114$&$0.128$\,\,\,&0.153&\nodata\\
\rownumber\  & \ SALT3.JLA&420&JLA&$0.142 \pm 0.011$&$2.952 \pm 0.124$&$0.145$\,\,\,&0.155&\,\,$0.032 \pm 0.028$\\
\\
&\ SALT2.JLA&420&K21 Valid&$0.151 \pm 0.008$&$3.064 \pm 0.086$&$0.103$\,\,\,&0.148&\nodata\\
\rownumber\  & \ SALT3.K21train&541 &K21 Valid&$0.115 \pm 0.008$&$2.967 \pm 0.091$&$0.118$\,\,\,&0.152&\,\,$0.013 \pm 0.020$\\
\\
&\ SALT2.JLA&420&K21 Full&$0.139 \pm 0.005$&$3.011 \pm 0.060$&$0.103$\,\,\,&0.146&\nodata\\
\rowcolor[gray]{0.9} \rownumber\  ~~~& \ SALT3.K21&1083&~~~~~K21 Full&~~~~~$0.140 \pm 0.005$&~~~~~$2.833 \pm 0.057$&~~~~~$0.110$\,\,\,&0.143&\,\,$-0.002 \pm 0.014$\\

        \hline\\*[-1.5ex]
    \multicolumn{9}{l}{
    \begin{minipage}{6.5in}
    
    Cosmological results from light-curve fits using SALT3 models created with different training sets, compared to results when using the SALT2.JLA model.  Training sets include a simulated JLA training sample (simJLA, Row 1), the real JLA training sample (JLA, Row 2), an expanded training sample using the full JLA training sample but including only half the additional data we use, with the remainder used for validation (K21train, Row 3), and the full K21 compilation (highlighted in Row 4), which is the training sample used to create the SALT3 model published in this work. Our best model using the complete K21 compilation results in nuisance parameters and distances statistically consistent but slightly lower Hubble scatter as compared to SALT2.
    
    $^{\rm a}$ Relative to the equivalent SALT2 fitting results, the distance between average Hubble residual at $0.01 < z < 0.2$ and the average Hubble residual at $0.4 < z < 0.6$.
    
    $^{\rm b}$ Large, combined simulations of CfA3, SDSS, and SNLS with a total of $\sim$3000~SNe to measure distance biases more precisely than is possible from a sample with the size of the JLA training sample
    
    \end{minipage}
    }
    \end{tabular}
    \label{table:saltstats}
\end{table*}

\section{Conclusions}
\label{sec:conclusions}

We have presented \texttt{SALTshaker}, a new Python-based training code to train a phenomenologically motivated light-curve model using the SALT framework, in addition to a retrained SALT model we call SALT3. \texttt{SALTshaker} is publicly available at \url{https://github.com/djones1040/SALTShaker} with documentation at  \url{https://saltshaker.readthedocs.io/en/latest/}. The \texttt{SALTshaker} documentation includes links to the training data and the SALT3.K21 model, and the SALT3.K21 model is compatible with and included in the latest versions of \texttt{SNANA} and \texttt{sncosmo}.  

The SALT3.K21 model itself includes updated calibration with Supercal, a training sample with 1083 SNe $-$ 2.5 times larger than previous training samples $-$ and extends to the rest-frame $iz$ bands. Due to its larger wavelength range, we find that SALT3.K21 distances for both legacy low-$z$ data and Foundation data are approximately 15\% more precise, equivalent to increasing the low-$z$ sample size by 30\%. The SALT3.K21 model is based on updated calibration with Supercal \citep{Scolnic2015} and revised MW E(B-V) estimates from \citet{Schlafly2011}. As part of an upcoming cosmology analysis we have employed SALT3.K21 within the \texttt{PIPPIN} \citep{pippin} framework, generating simulations and performing bias corrections following the methodologies of \cite{Kessler18} and \cite{KesslerScolnic2017}. 
Although this work is preliminary, we find that cosmological parameters found using a larger light-curve dataset of $\sim 2000$ \sne\ (to be released in Brout+ in prep) are consistent between SALT3.K21 and SALT2.JLA. Looking to future missions, we find that for the forecast \sn\ survey of the \textit{Roman Space Telescope}, assuming the ALL$z$ strategy \citep{Hounsell2018}, the extended wavelength range in SALT3.K21 makes use of $\sim 20\%$ more observations compared to SALT2.JLA; this increase is about a factor of 2 for redshifts $ z<0.5$, and falls to about a 10\% increase at $z=1.5$.

Several light-curve models have been developed for cosmological supernovae, including MLCS \citep{Riess1996MLCS}, MLCS2k2 \citep{Jha2007}, SiFTO \citep{Conley2008}, SNooPy \citep{Burns2011}, SNEMO \citep{Saunders2018}, SUGAR \citep{Leget2019}, and BayesSN \citep{Mandel2011,Mandel2020}. In the context of other modern light-curve models, SALT3 offers an approach to model design and training process that prioritizes the use of heterogeneous spectral and photometric data to provide extensive phase and wavelength coverage and native k-corrections through cosmology-independent training. The model framework is minimally changed from SALT2, so  \texttt{SNANA} simulations, bias corrections, and other analysis products are expected to require little revision.

Over the coming years, we expect \texttt{SALTshaker} will continue to be developed and improved as additional SN data becomes available and additional SN standardization parameters (e.g., host mass) are discovered and explored. Further development work could focus on the error model, which is currently based on central filter wavelengths rather than integrated quantities. This is a potential source of systematic uncertainty because observer-frame filter functions are contracted in the rest frame. 
Additionally, \texttt{SALTshaker} enables a more rigorous evaluation of systematic uncertainties such as those arising from limited training data, photometric calibration uncertainties, or treatment of SN spectra. These can be evaluated in a straightforward and rigorous way by re-training the SALT3 model surfaces on simulated data. 
Although we have demonstrated that \texttt{SALTshaker} can faithfully recover a truth model at the $\sim$1\% level, future work will also be able to fully validate the model training process using an entire analysis chain that includes training, bias corrections, and cosmology fitting  (Dai et al in prep). Similarly, while the SALT3 model surfaces presented in this work have been trained on data recalibrated to the level of $1\%$ via the Supercal procedure \citep{Scolnic2015,Scolnic2018}, we have left quantifying the reduced calibration uncertainties as a topic for further work. 

The \texttt{BYOSED} code \citep{Pierel2020} implements a range of simulated effects to a base SED template, such as perturbations to line velocity, multiple sources of reddening with distinct effects, and correlations of host galaxy properties with the \sn\ SED. Future work could perform the \texttt{SALTshaker} method on a \texttt{BYOSED}-produced training sample with such underlying effects.  By propagating any biases introduced by the training code into cosmology, we may quantify the potential impact of currently unmodeled supernova phenomenology on cosmology.

Samples of \sn\ light curves will increase by orders of magnitude with the Vera Rubin Telescope's LSST \citep{Ivezic2019} and the Roman Space Telescope \citep{Hounsell2018}. For error budgets to continue improvement, light-curve models should not be tied to outdated calibration standards, and it is essential that the model training process be regarded as a key component of an integrated cosmology analysis, as has been done in \citet{Betoule2014} and \citet{Mosher2014}.


\software{AstroPy \citep{astropy2013,astropy2018},Astroquery \citep{Ginsburg2019astroquery},extinction\citep{barbary2016extinction},iMinuit \citep{James1975,dembinski2020},
Matplotlib \citep{Hunter2007matplotlib}, NumPy \citep{harris2020array},SciPy \citep{2020SciPy-NMeth}, sncosmo \citep{Barbady2015sncosmo,Barbady2016sncosmo}, tqdm \citep{costa_luis_2021_tqdm}
}

\acknowledgements

We thank Julien Guy and the developers of the SALT and SALT2 models for their contributions to the field, Julien Guy, Chris Lidman, and Georgie Taylor for valuable assistance and discussion, and  Julien Guy, Marc Betoule, and the SNLS team for providing the original \texttt{snpca} code and JLA training sample. We are grateful for all astronomers who acquired the photometry and spectroscopy used to train the SALT3.K21 model.

D.O.J. is supported by a Gordon and Betty Moore Foundation postdoctoral fellowship at the University of California, Santa Cruz and by NASA through the NASA Hubble Fellowship grant HF2-51462.001 awarded by the Space Telescope Science Institute, which is operated by the Association of Universities for Research in Astronomy, Inc., for NASA, under contract NAS5-26555. Support for this work was provided by the STScI Director’s Discretionary Fund. This research at Rutgers University (S.W.J., M.D.) was supported by NASA contract NNG16PJ34C and DOE award DE-SC0011636. M.D. is also supported by the Horizon Fellowship at the Johns Hopkins University. This work at the University of South Carolina (S.~Rodney, J.~D.~R.~Pierel) was supported by grant HST-AR-15808 from the Space Telescope Science Institute.  This work was supported by NASA contract No.\ NNG17PX03C issued through the Nancy G.\ Roman Science Investigation Teams Program.

This research is based on observations made with the NASA/ESA Hubble Space Telescope obtained from the Space Telescope Science Institute, which is operated by the Association of Universities for Research in Astronomy, Inc., under NASA contract NAS 5–26555. These observations are associated with programs GO-4016, GO-4252, GO-8611, GO-9114, and GO-10182.

This research is based on observations made with the International Ultraviolet Explorer, obtained from the MAST data archive at the Space Telescope Science Institute, which is operated by the Association of Universities for Research in Astronomy, Inc., under NASA contract NAS 5–26555. These observations are associated with programs METOO, SNMRK, SNNRK, STKRK, and VILSP.

This paper is based in part on observations made with the Southern African Large Telescope (SALT) using the Robert Stobie Spectrograph (RSS), through allocations made to Rutgers University via programs 2015-1-MLT-002, 2016-1-MLT-007, and 2017-1-MLT-002 (PI: S.W.\ Jha).

Based in part on observations obtained at the Southern Astrophysical Research (SOAR) telescope (NOIRLab Prop.\ IDs 2015A-0253, 2015B-0313, 2017A-0306, 2017B-0169, 2018A-0277; PI: R.\ Foley), which is a joint project of the Minist\'erio da Ci\^encia, Tecnologia e Inova\c{c}\~oes do Brasil (MCTI/LNA), the US National Science Foundation's NOIRLab, the University of North Carolina at Chapel Hill (UNC), and Michigan State University (MSU).  

Based in part on observations at Kitt Peak National Observatory at NSF's NOIRLab (NOIRLab Prop.\ IDs 2015A-0253, 2015B-0313, 2017A-0306, 2017B-0169, 2018A-0277; PI: R.\ Foley), which is managed by the Association of Universities for Research in Astronomy (AURA) under a cooperative agreement with the National Science Foundation. The authors are honored to be permitted to conduct astronomical research on Iolkam Du'ag (Kitt Peak), a mountain with particular significance to the Tohono O'odham.

A major upgrade of the Kast spectrograph on the Shane 3 m telescope at Lick Observatory was made possible through generous gifts from the Heising-Simons Foundation as well as William and Marina Kast.  Research at Lick Observatory is partially supported by a generous gift from Google. 

Some of the data presented herein were obtained at the W.\ M.\ Keck Observatory, which is operated as a scientific partnership among the California Institute of Technology, the University of California and the National Aeronautics and Space Administration. The Observatory was made possible by the generous financial support of the W.\ M.\ Keck Foundation.  The authors wish to recognize and acknowledge the very significant cultural role and reverence that the summit of Maunakea has always had within the indigenous Hawaiian community.  We are most fortunate to have the opportunity to conduct observations from this mountain.

\bibliographystyle{apj}
\bibliography{references}

\end{document}